\def\ifhtx{\iffalse}    
\def\editrevision{2}\def\showrevision{0} 
\long\def\revision#1#2#3{\ifnum#1>\editrevision{#3}\else %
  \ifnum\showrevision>1{\em\fg{DarkGoldenrod}[#3]}\fi
  \ifnum\showrevision>0{\fg{blue}#2}\else{#2}\fi
  \fi}
\ifhtx                  
\def\Arefx#1#2{\Aref{#1}{#2}}
\def\Arefs#1#2{\Aref{#1}{#2}}
\def\Tref#1#2{\Aref{#1}{#2}}
\def\Fref#1#2{\Aref{#1}{#2}}
\def\twikiname#1#2{\makebox[10em][l]{\qquad\A{http://www.ivoa.net/twiki/bin/view/IVOA/#1}{#2}}}

\def\texorpdfstring#1#2{#1}
\cgidef{ -tex
\filex  html
\slash  {{\fg{blue}/}}
\attr   #1{{\tt{\fg{DarkRed}#1}}}
\elem   #1{{\tt{\fg{DarkRed}#1}}}
\elemdef #2{{\fg{blue}$<$}{\tt{\fg{DarkRed}#1}#2}{\fg{blue}$>$}}
\xtag    #1{\elemdef{#1}{}}
\attrval #2{{\tt{\fg{DarkRed}#1}="{\fg{DarkMagenta}#2}"}}
\elemdef #2{{\fg{blue}<}{\tt{\fg{DarkRed}#1}#2}{\fg{blue}>}}
\literalvalue   #1{"{\fg{DarkMagenta}#1}"}
{comment} #1{{\fg{darkblue}\em(#1)} }{}
\Plain #1{\plain{#1}}
\order  {~\image{oplus.gif} }
\unorder        {~\image{circ2.gif} }
\deprecated     {~\image{dagger.gif} }
\choice   {\image{leadsto2.gif} }
\left   #1{}
\right  #1{}
\requiredattr #1{{\tt\bf{\fg{DarkBlue}#1}}}
\modified #3 {#2}  
}
\else                   
\documentclass[10pt,notitlepage,onecolumn,lcustom]{ivoa}
\usepackage{color}              
\usepackage{verbatim}           

\ifx\pdftexversion\undefined 
  \usepackage[dvips]{graphicx}
  \DeclareGraphicsExtensions{.eps,.ps}
  \usepackage[ps2pdf]{hyperref}

\else                        
  \usepackage[pdftex]{graphicx} 
  \usepackage{epstopdf}         
  \makeatletter
  \g@addto@macro\Gin@extensions{,.ps}
  \@namedef{Gin@rule@.ps}#1{{pdf}{.pdf}{`ps2pdf #1}}
  \makeatother
  \usepackage[pdftex,bookmarks=true,bookmarksnumbered=true,hypertexnames=false,breaklinks=true,%
  colorlinks,linkcolor=blue,urlcolor=blue]{hyperref}
  \let\A=\href
  \def\Aref#1#2{section~\ref{#1}}
  \def\Arefs#1#2{section~\ref{#1}}
  \def\Arefx#1#2{appendix~\ref{#1}}
  \def\Tref#1#2{Table~\ref{#1}}
  \def\Fref#1#2{Figure~\ref{#1}}
\fi

\let\fg=\color                  
\let\Beg=\begin
\definecolor{DarkRed}{rgb}{0.5,0,0}
\definecolor{DarkBlue}{rgb}{0,0,0.5}
\definecolor{DarkGreen}{rgb}{0,0.5,0}
\definecolor{DarkPurple}{rgb}{0.3,0.1,0.5}
\definecolor{DarkGoldenrod}{rgb}{0.72,0.5,0.05}
\def\modified#1#2#3{#2}  
\def\filex{pdf} 
\def\slash {{\fg{blue}/}}
\def\attr#1{{\tt{\fg{DarkRed}#1}}}
\def\requiredattr#1{{\tt\bf{\fg{DarkBlue}#1}}}
\def\elem#1{{\tt{\fg{DarkRed}#1}}}
\def\attrval#1#2{{\tt{\fg{DarkRed}#1}="{\fg{DarkPurple}#2}"}}
\def\elemdef#1#2{{\fg{blue}$<$}{\tt{\fg{DarkRed}#1}#2}{\fg{blue}$>$}}
\def\literalvalue#1{{\tt"}{\fg{DarkPurple}#1}{\tt"}}
\def\order{$\oplus$ }
\def\unorder{{\large $\circ$ }}
\def\deprecated {$\dagger$ }
\def\choice{{$\mapsto$ }}
\def\twikiname#1#2{\makebox[10em][l]{{#2}}}
\def\thickrule{\noindent\rule{\textwidth}{1pt}}
\def\Plain#1{{\sf #1}}
\begin{document}
\newenvironment{TABULAR}[2]{\begin{tabular}{#2}}{\end{tabular}}
\newenvironment{plain}{\begin{quote}}{\end{quote}}
\fi


\title{VOTable Format Definition}
\date{2013-09-20}

\ivoatype{IVOA Recommendation}

\version{1.3}

\editor{Fran\c cois {Ochsenbein}\\
        \hspace*{1.7em}
        Mark Taylor}
\urlthisversion{
  \url{http://www.ivoa.net/Documents/VOTable/20130920/}} 
\urllastversion{
  \url{http://www.ivoa.net/Documents/latest/VOT.html}}
\previousversion{
  \url{http://www.ivoa.net/Documents/VOTable/20091130/}
  \quad V1.2 (2009-11-30) \hfill\\
  \hspace*{1.8em}
  \url{http://www.ivoa.net/Documents/cover/VOT-20040811.html}
  \quad V1.1 (2004-08-11) \hfill\\
  \hspace*{1.8em}
  \url{http://www.ivoa.net/Documents/PR/VOTable/VOTable-20031017.html}
  \quad V1.0 (2002-04-15) 
}
\author{
\normalsize \twikiname{FrancoisOchsenbein}{Fran\c cois {\bf Ochsenbein}}
        \quad{\em Observatoire Astronomique de Strasbourg, France} \\
\normalsize \twikiname{RoyWilliams}{Roy {\bf Williams}}
        \quad{\em California Institute of Technology, USA} \\
\normalsize\ifhtx\else\hspace*{-0.75em}\fi 
   {{\em with contributions from:}}\\
\normalsize \twikiname{CliveDavenhall}{Clive {\bf Davenhall}}
        \quad{\em University of Edinburgh, UK} \\
\normalsize \twikiname{MarkusDemleitner}{Markus {\bf Demleitner}}
        \quad{\em Heidelberg University, Germany} \\
\normalsize \twikiname{CanielDurand}{Daniel {\bf Durand}}
        \quad{\em Canadian Astronomy Data Centre, Canada} \\
\normalsize \twikiname{PierreFernique}{Pierre {\bf Fernique}}
        \quad{\em Observatoire Astronomique de Strasbourg, France} \\
\normalsize \twikiname{DavidGiaretta}{David {\bf Giaretta}}
        \quad{\em Rutherford Appleton Laboratory, UK} \\
\normalsize \twikiname{BobHanisch}{Robert {\bf Hanisch}}
        \quad{\em Space Telescope Science Institute, USA} \\
\normalsize \twikiname{TomMcGlynn}{Tom {\bf McGlynn}}
        \quad{\em NASA Goddard Space Flight Center, USA} \\
\normalsize \twikiname{AlexSzalay}{Alex {\bf Szalay}}
        \quad{\em Johns Hopkins University, USA} \\
\normalsize \twikiname{MarkTaylor}{Mark {\bf Taylor}}
        \quad{\em University of Bristol, UK} \\
\normalsize \twikiname{AndreasWicenec}{Andreas {\bf Wicenec}}
        \quad{\em European Southern Observatory, Germany} \\
}


\maketitle 
  
\section*{Abstract}
This document describes the structures making up
the VOTable standard.

\noindent The main part of this document describes the adopted part of the
VOTable standard; it is followed by appendices presenting extensions
which have been proposed and/or discussed, but which are not part of 
the standard.

\section*{Status of This Document}

This document has been produced by the IVOA Applications Working Group,
building on the work of the currently dormant IVOA VOTable Working Group.

It has been reviewed by IVOA Members and other interested parties,
and has been endorsed by the IVOA Executive Committee as an IVOA
Recommendation.
It is a stable document and may be used as reference material or cited as
a normative reference from another document.  IVOA's role in making the
Recommendation is to draw attention to the specification and to promote
its widespread deployment.  This enhances the functionality and
interoperability inside the Astronomical Community.

\clearpage
\ifhtx\Beg{tabular}{cellpadding=5 \bg{LightYellow}}{||p||}\fi
\tableofcontents
\ifhtx\End{tabular}\fi

\clearpage
\section{Introduction}

The VOTable format is an XML standard for the interchange of data
represented as a set of tables. 
In this context, a table is an unordered set of rows, each of
a uniform structure, as specified in the table description
(the table {\em metadata}). 
Each row in a table is a sequence of table cells, and each of these contains
either a primitive data type, or an array of such primitives. 
VOTable is derived from the
Astrores format [1], itself modeled on the FITS Table format [2];
VOTable was designed to be close to the FITS Binary Table format.

\subsection{Why VOTable?}

Astronomers have always been at the forefront of developments in
information technology, and funding agencies across the world have
recognized this by supporting the Virtual Observatory movement, in
the hopes that other sciences and business can follow their lead in
making online data both {\it interoperable} and {\it scalable}.

VOTable is designed as a flexible storage and exchange format for
tabular data, with particular emphasis on astronomical tables.

Interoperability is encouraged through the use of standards (XML).
The XML fabric
allows applications to easily validate an input document, as well as
facilitating transformations through XSLT (eXtensible Style Language
Transformation) engines.

\subsubsection*{Grid Computing}

VOTable has built-in features for big-data and Grid computing. It
allows metadata and data to be stored separately, with the remote
data linked. 
Processes can then use
metadata to `get ready' for their input data, or to organize
third-party or parallel transfers of the data. Remote data allow the
metadata to be sent in email and referenced in documents without
pulling the whole dataset with it: just as we are used to the idea of
sending a pointer to a document (URL) in place of the document, so we
can now send metadata-rich pointers to data tables in place of the
tables themselves. The remote data is referenced with the URL syntax
{{\sf protocol://location}},
meaning that arbitrarily complex protocols are allowed.

When we are working with very large tables in a
distributed-computing environment (``the Grid"), the data
stream between processors, with flows being filtered, joined, and
cached in different geographic locations. It would be very difficult
if the number of rows of the table were required in the header --
we would need to stream in the whole table into a cache, compute the
number of rows, then stream it again for the computation. In the
Grid-data environment, the component in short supply is not the
computers, but rather these very large caches. Furthermore, these
remote data streams may be created dynamically by another process or
cached in temporary storage: for this reason VOTable can express that
remote data may not be available after a certain time (\attr{expires}).
Data on the net may require authentication for access, so VOTable
allows expression of password or other identity information (the
`{\attr{rights}}'
attribute).

\subsubsection*{Data Storage: Flexible and Efficient}

The data part in a  VOTable may be represented using one of four
different formats: TABLEDATA, FITS, BINARY and BINARY2. TABLEDATA is a
pure XML format so that small tables can be easily handled in their
entirety by XML tools. The FITS binary table format is well-known to
astronomers, and VOTable can be used either to encapsulate such a
file, or to re-encode the metadata; unfortunately it is difficult to
stream FITS, since the dataset size is required in the header 
(NAXIS2 keyword), and FITS requires a specification up front of the maximum
size of its variable-length arrays. The BINARY and BINARY2 formats
are supported for efficiency and ease of programming: no FITS
library is required, and the streaming paradigm is supported.

VOTable can be used in different ways, as a data
storage and transport format, and also as a way to store metadata
alone (table structure only).  In the latter case, a
VOTable structure can be sent to a server, which can then open a
high-bandwidth connection to receive the actual data, using the
previously-digested structure as a way to interpret the stream of
bytes from the data socket. 

VOTable can be used for small numbers of small records (pure XML
tables), or for large numbers of simple records (streaming data), or
it can be used for small numbers of larger objects. In the latter
case, there will be software to spread large data blocks among
multiple processors on the Grid. Currently the most complex structure
that can be in a VOTable Cell is a multidimensional array.

\subsection{XML Conventions}

VOTable is constructed with \A{http://www.w3.org/XML/}{XML} (extensible Markup Language), a
powerful standard for structured data throughout the Internet
industries. It derives 
from SGML, 
a standard used in the publishing industry and for 
technical documentation for many years. XML
consists of {\it elements} and payload, where an element consists of
a {\it start tag} (the part in angle brackets), the payload, and an
{\it end tag} (with angle brackets and a slash). Elements can
contain other elements. Elements can also bear
{\attr{attributes}}
(keyword-value combinations).

The payload may be in two forms: parsed or unparsed character
data. Examples are:

\begin{verbatim}
<text>Fran&#231;ois</text>
<text><![CDATA[ a & (b <= c) ]]></text>
\end{verbatim}

In the first example, the sequence {\tt \&\#231;} is interpreted as
part of the ISO/IEC 10646 character set (Unicode), and translates to an
accented character, so that the text is ``Fran\c{c}ois".
The second example uses the special {\tt CDATA} sequence so that the
characters {\tt <}, {\tt >}, and {\tt\&} can be used without interpretation;
in this case, any ASCII characters are allowed except the terminating
sequence {\tt]]>}. For more information, see any book on
XML.

\subsection{Syntax Policy}

Following the general XML rule, element and attribute names are
case-sensitive and have to be used with the specified 
capitalisation. For VOTable, we have adopted the convention that
element names are spelled in uppercase
and attribute names in lowercase (with an
exception for the {\attr{ID}}
attribute). 
Element and attribute names are further distinguished in
this paper by being typed with a {\attr{red fixed-width}} font,
and the values of the attributes by being \literalvalue{coloured}.

\subsection{VOTable in the VO Architecture}
\label{sec:voarch}

\ifhtx\begin{tabular}{c}
\tag{IMG SRC="ivoa-archi.png" ALT="VOTable in IVOA Architecture"
     ALIGN="LEFT" BORDER="0"}
\end{tabular}
\else
\begin{figure}[h]
\includegraphics[width=\textwidth]{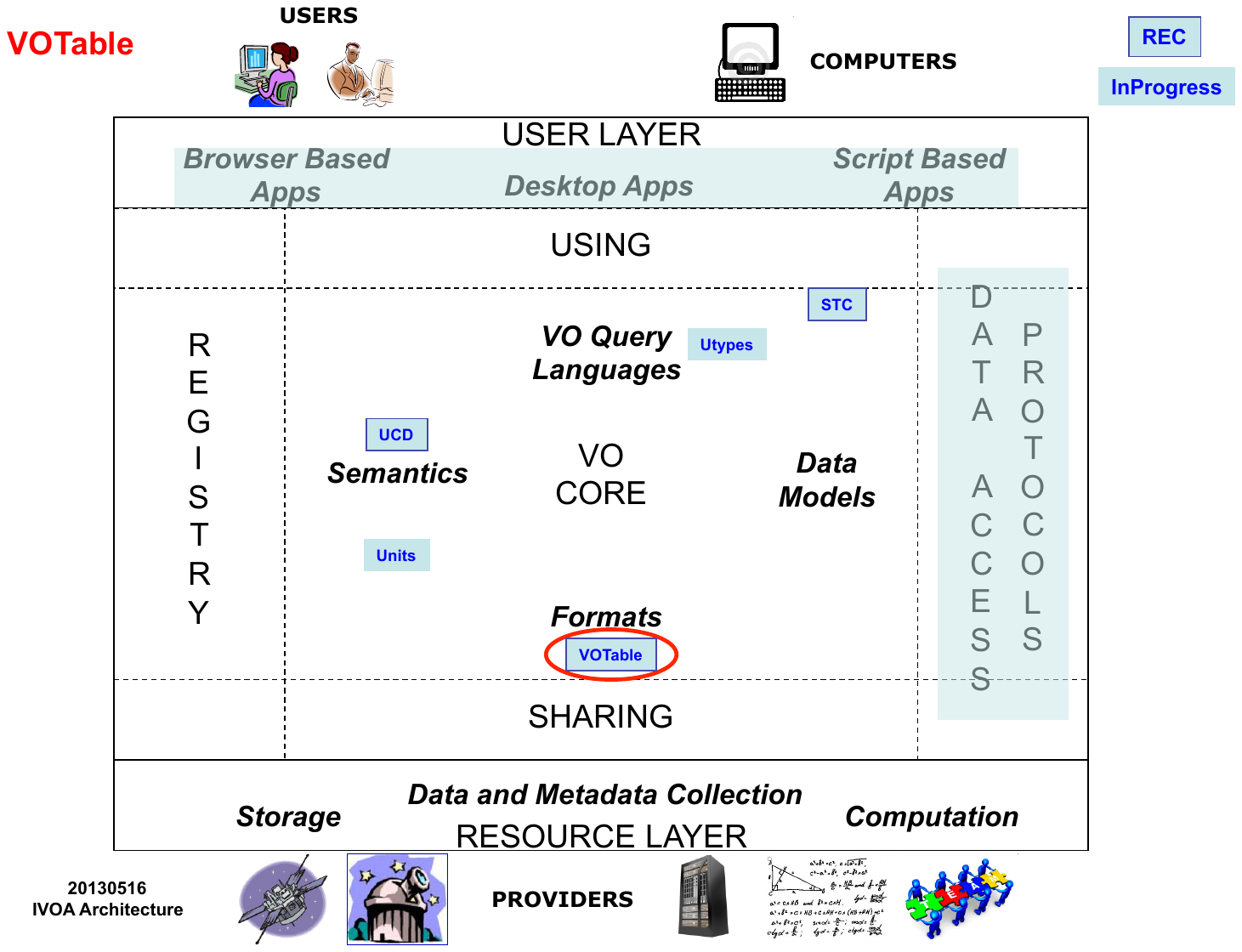}
\caption{\label{fig:vo-arch}VOTable in the IVOA Architecture}
\end{figure}
\fi

VOTable is a core IVOA standard.

Wherever tabular data is transferred between Virtual Observatory components,
VOTable provides the preferred serialization format.
Since tables are used to list available resources as well as to
represent science data which is itself tabular,
this means that VOTable is used pervasively in the definitions
of the Data Access protocols (e.g.\ SCS, SIA, SSA, TAP),
and hence for exchange of data and metadata
between user layer applications and data-providing services.
VOTable is also used as a serialization format for
some of the IVOA Data Models.

In order to represent semantically rich metadata, VOTable relies on
the other IVOA standards UCD, Utype, Units and STC.
This document explains how information structured according to those
standards are managed within the VOTable framework.

\section{Data Model}

In this section we define the data model of a VOTable, and in the
next sections its syntax when expressed as XML. The data model of
VOTable can be expressed as:    

\medskip
\ifhtx\begin{tabular}{RRCp{0.8\textwidth}}
\else\begin{tabular}{rrcp{0.7\textwidth}}\fi
\hspace{3em}&{\bf VOTable} &=& hierarchy of {\bf Metadata} + associated
        {\bf TableData}, arranged as a set of {\bf Tables}\\
&{\bf Metadata} &=& {\bf Parameters} + {\bf Infos} + {\bf Descriptions}
                + {\bf {\fg{black}Links + Fields + Groups}}\\
&{\bf Table} &=& list of {\bf Fields + TableData}\\
&{\bf TableData}{ } &=& stream of {\bf Rows}\\
&{\bf Row} &=& list of {\bf Cells}\\
&{\bf Cell} &=& 
        $\left\{
        \begin{tabular}{l}
         {\bf Primitive} \\
        or variable-length list of {\bf Primitives} \\
        or multidimensional array of {\bf Primitives}\\
        \end{tabular}
        \right.$
        \\
&{\bf Primitive} &=& integer, character, float, floatComplex, etc
(see \Tref{primitives}{table of primitives} below).
\end{tabular}

\medskip
\par\noindent
Metadata is divided into that which concerns the table itself 
(parameters), and the definitions of the fields (or column
attributes) of the table. 
Each \elem{FIELD} represents the metadata 
that can be found at the
top of the column in a paper version of the table: 
in the example introduced in \Aref{example1}{the
section} below, the first \elem{FIELD} has its \attr{name} attribute
set to \literalvalue{RA}. The Field can be thought of as a class definition,
and the table cells below it are the instances of that class.

A parameter ({\elem{PARAM}})
is similar to a {\elem{FIELD}},
except that it has a \attr{value} attribute.
Parameters can be seen as ``constant columns'', containing for instance
FITS keywords or any other
information pertaining to the table itself or its environment, such as the
{\tt Telescope} parameter in the example of \Arefs{example1}{section 3.1}.

An informative parameter ({\elem{INFO}}) (see \Arefs{elem:INFO}{INFO})
is a restricted form of the {\elem{PARAM}} ---  it is always understood
as a {\em string} (i.e. \attrval{datatype}{char}
and \attrval{arraysize}{*} are {\em implied}).

\mbox{}{\color{black}}%
The ordered list of Fields at the top of the table thus provides a
template for a Row object (also called a {\it record}). The
template allows interpretation of the data in the Row. 
The
record is a set of Cells, with the number and order of Cells the same for each
Row, and the same as the number of Fields defined in the Metadata.

In VOTable,
there is generally no advance specification of the number of rows in the table:
this is to allow streaming of large tables, as discussed above.
However, if the number of rows is known, it may be specified in a
dedicated \attr{nrows} attribute.

From Version 1.1, columns may be logically grouped, so that it is
possible to define table substructures made of column associations.
Such an association is declared as a \elem{GROUP}, which typically
contains column references (\elem{FIELDref}) 
and associated parameters (\elem{PARAM}).

\subsection{Primitives}

\ifhtx\label{primitives}
\begin{center}\Beg{tabular}{CELLSPACING=4 CELLPADDING=4}{|rlrl|}
\else
\begin{table}[hbt]
 \begin{center}\begin{tabular}{|r|l|c|r|}
\fi\hline
  {\attr{datatype}} & Meaning & \attr{FITS} &
      { Bytes} \\
 \hline
 \literalvalue{boolean}      & Logical         &\literalvalue{L}& 1  \\
 \literalvalue{bit}          & Bit             &\literalvalue{X}& *  \\
 \literalvalue{unsignedByte} & Byte (0 to 255) &\literalvalue{B}& 1  \\
 \literalvalue{short}        & Short Integer   &\literalvalue{I}& 2  \\
 \literalvalue{int}          & Integer         &\literalvalue{J}& 4  \\
 \literalvalue{long}         & Long integer    &\literalvalue{K}& 8  \\
 \literalvalue{char}         & ASCII Character &\literalvalue{A}& 1  \\
 \literalvalue{unicodeChar}  & Unicode Character&        & 2 \\
 \literalvalue{float}        & Floating point  &\literalvalue{E}& 4  \\
 \literalvalue{double}       & Double          &\literalvalue{D}& 8  \\
 \literalvalue{floatComplex} & Float Complex   &\literalvalue{C}& 8  \\
 \literalvalue{doubleComplex}& Double Complex  &\literalvalue{M}& 16 \\
\hline\end{tabular}\end{center}
\ifhtx\par
\else\caption{\label{primitives}List of the Primitives
{\em(details in \Aref{sec:datatypes}{})}}\end{table}
\fi

Each Cell is composed from Primitives, each of which is a datatype
of fixed-length binary representation, as listed in 
\Tref{primitives}{the accompanying table}.
\modified{1.2-090606}{
Cells may consist of a single Primitive (this is
the default), or of an {\em array} (which may be multidimensional)
of Primitives (see \Aref{array}{the next section}).}{
Cells may consist of a single Primitive (this is
the default), or of a multidimensional array of Primitives (see
\Aref{array}{the next section}).}

Except for the Bit type, each primitive has the fixed length in
bytes given in \Tref{primitives}{the table}. 
Bit scalars and arrays are stored in
the minimum number of bytes feasible (so that $b$ bits take the integer
part of $(b+7)/8$ bytes).  These primitives
are described in more detail in \Aref{sec:datatypes}{section 6}.

VOTables support two kinds of characters: ASCII 1-byte characters
and Unicode (UCS-2) 2-byte characters. Unicode is a way to represent
characters that is an alternative to ASCII. It uses two bytes per
character instead of one, it is strongly supported by XML tools, and
it can handle a large variety of international alphabets. Therefore
VOTable supports not only ASCII strings ({\attrval{datatype}{char}}),
but also Unicode ({\attrval{datatype}{unicodeChar}}).

Note that strings are not a primitive type: strings are 
represented in VOTable as an array of characters. 

\modified{1.2-090606}{
\subsection{Columns as Arrays}\label{array}}{
\subsection{Multidimensional Arrays}\label{array}}
\label{sec:dim}

A table cell can contain an {\em array} of a given primitive type,
with a fixed or variable number of elements; the array may even
be multidimensional. For instance, the position of a point in a
3D space can be defined by the following:

\elemdef{FIELD}{ \attrval{ID}{point\_3D} \attrval{datatype}{double}
   \attrval{arraysize}{3}\slash}

\noindent and each cell corresponding to that definition must contain exactly
3 numbers. An asterisk ({\bf\tt*}) may be appended to indicate
a {\em variable} number of elements in the array, as in:

\elemdef{FIELD}{ \attrval{ID}{values} \attrval{datatype}{int}
   \attrval{arraysize}{100*}\slash}

\noindent where it is specified that each cell corresponding to that 
  definition contains 0 to 100 integer numbers. The number may be 
  omitted to specify an unbounded array
  (in practice up to $\simeq 2\times10^9$ elements).

A table cell can also contain a {\em multidimensional array} 
of a given primitive type. This is specified by a sequence of dimensions
separated by the {\tt x} character, 
with the first dimension changing fastest; as in the case
of a simple array, the last dimension may be variable in length.
As an example, the following definition
declares a table cell which may contain a set of up to 10 images,
each of 64x64 bytes:

\elemdef{FIELD}{ \attrval{ID}{thumbs} \attrval{datatype}{unsignedByte} 
  \attrval{arraysize}{64x64x10*}\slash}

Strings, which are defined as a set of characters,
can therefore be represented in VOTable as a fixed- or variable-length 
array of characters:

\elemdef{FIELD}{ \attrval{name}{unboundedString} \attrval{datatype}{char}
       \attrval{arraysize}{*}\slash}

A 1D array of strings can be represented as a 2D array of characters, but
given the logic above, it is possible to define a variable-length array
of fixed-length strings,
but not a fixed-length array of variable-length strings.
A convention to express an array of variable-length strings
exists (see \Aref{sec:arraystring}{in the appendix}) but is not
part of this standard.

\subsection{Compatibility with FITS Binary Tables}

VOTable is closely compatible with the FITS Binary Table format.
Henceforth, we shall abbreviate ``FITS Binary Table  and its
Conventions" simply by the word ``FITS". Given a FITS
file that represents a binary table, the header may be converted to
VOTable, with a pointer to the original file, or with the original
file included directly in VOTable. Since the original file is still
present, it is clear that no data has been lost. A {\elem{PARAM}}
element can be used to hold any FITS keyword with its value
and comment string.

We might ask two more significant questions, about how much of
the FITS header and data can be represented in VOTable. The answer is
that there is considerable overlap. 

For instance, the recommended formatting of the data for an
edition of the data is expressed by the non-mandatory TDISP keyword:
for example F12.4 means 12 characters are to be used, and 4 decimal
places. This has been converted in VOTable as the attributes {\attr{width}}
and {\attr{precision}}
which, connected with {\bf {\attr{datatype}}},
are semantically identical to the TDISP keyword. 

\subsubsection*{What can FITS do but not VOTable?}

FITS has complex semantics, with many conventions
(see {\em e.g.} the \A{http://fits.gsfc.nasa.gov/fits_registry.html}
{Registry of FITS Conventions} [10]) which have been developed
mainly to be able to cope with the increasing complexity
of astronomical instrumentation. In the frame of the
{\em Virtual Observatory} the complexity is described by
means of {\em data models}, and from its version 1.1,
{\em VOTable} can refer to these data models by means
of the \attr{utype} attribute described in
\Aref{sec:utype}{section 4.6}.

\subsubsection*{What can VOTable do but not FITS?}

VOTable supports separating of data from metadata and the
streaming of tables, and other ideas from modern distributed
computing. It bridges two ways to express structured data: XML and
FITS. It uses UCDs -- see \Aref{sec:ucd}{below}) 
to formally express the semantic
content of a parameter or field. It has the hierarchy and flexibility
of XML: using \elem{GROUP} elements introduced in version 1.1, 
columns in a VOTable can be grouped in arbitrarily complex hierarchies; 
and the ID attribute can be used in XML 
to enable what are essentially pointers.
FITS does not handle Unicode (extended alphabet) characters.

\medskip

\noindent{\fg{black}It should be noticed that the transformation
of FITS to VOTable is reversible: 
any FITS table can be converted to a VOTable without loss of
information and the resulting VOTable can be converted back to a
FITS table also without loss of information.
However, it is
possible to create new VOTables which cannot be converted to FITS
tables without loss of information.
}

\section{The VOTable Document Structure}
\label{elem:VOTABLE}

The overall VOTable document structure is described and controlled
by its \A{http://www.ivoa.net/xml/VOTable/v1.3}{XML Schema}.
That means that documents claiming to represent VOTables must
include the reference to the VOTable schema, and 
pass through W3C XML Schema validators without error;
notice that the validation is a necessary, {\em but not sufficient}, 
condition for correctness.
The XML Schema of this version 1.3 is included in 
\Arefx{XML-schema}{Appendix B},
and is illustrated in \Arefs{dtd}{section 7}.

A
 VOTable document consists of a single all-containing element
called {\elem{VOTABLE}},
which contains descriptive elements and global definitions
({\elem{DESCRIPTION}}, \elem{GROUP}, \elem{PARAM}, \elem{INFO}),
followed by one or more {\elem{RESOURCE}} elements.
Each Resource element contains zero or more \elem{TABLE} elements,
and possibly other \elem{RESOURCE} elements. 

The \elem{TABLE} element, the actual heart of VOTable, contains 
a description of the columns and parameters 
(described in \Aref{sec:field}{the next section})
followed by the data values 
(described in \Aref{sec:data}{the following section}).

\subsection{Example}

This simple example of a VOTable document lists 3 galaxies with their
position, velocity and error, and their estimated distance.

\label{example1}
\ifhtx\Beg{tabular}{\bg{LightCyan} CELLPADDING=5}{||l||}
\else\begingroup\small
\fi
\begin{verbatim}
<?xml version="1.0"?>
<VOTABLE version="1.3" xmlns:xsi="http://www.w3.org/2001/XMLSchema-instance"
 xmlns="http://www.ivoa.net/xml/VOTable/v1.3"
 xmlns:stc="http://www.ivoa.net/xml/STC/v1.30" >
  <RESOURCE name="myFavouriteGalaxies">
    <TABLE name="results">
      <DESCRIPTION>Velocities and Distance estimations</DESCRIPTION>
      <GROUP utype="stc:CatalogEntryLocation">
        <PARAM name="href" datatype="char" arraysize="*"
               utype="stc:AstroCoordSystem.href" value="ivo://STClib/CoordSys#UTC-ICRS-TOPO"/>
        <PARAM name="URI" datatype="char" arraysize="*"
               utype="stc:DataModel.URI" value="http://www.ivoa.net/xml/STC/stc-v1.30.xsd"/>
        <FIELDref utype="stc:AstroCoords.Position2D.Value2.C1" ref="col1"/>
        <FIELDref utype="stc:AstroCoords.Position2D.Value2.C2" ref="col2"/>
      </GROUP>
      <PARAM name="Telescope" datatype="float" ucd="phys.size;instr.tel"
             unit="m" value="3.6"/>
      <FIELD name="RA"   ID="col1" ucd="pos.eq.ra;meta.main"
             datatype="float" width="6" precision="2" unit="deg"/>
      <FIELD name="Dec"  ID="col2" ucd="pos.eq.dec;meta.main"
             datatype="float" width="6" precision="2" unit="deg"/>
      <FIELD name="Name" ID="col3" ucd="meta.id;meta.main"
             datatype="char" arraysize="8*"/>
      <FIELD name="RVel" ID="col4" ucd="spect.dopplerVeloc" datatype="int"
             width="5" unit="km/s"/>
      <FIELD name="e_RVel" ID="col5" ucd="stat.error;spect.dopplerVeloc"
             datatype="int" width="3" unit="km/s"/>
      <FIELD name="R" ID="col6" ucd="pos.distance;pos.heliocentric"
             datatype="float" width="4" precision="1" unit="Mpc">
        <DESCRIPTION>Distance of Galaxy, assuming H=75km/s/Mpc</DESCRIPTION>
      </FIELD>
      <DATA>
        <TABLEDATA>
        <TR>
          <TD>010.68</TD><TD>+41.27</TD><TD>N 224</TD><TD>-297</TD><TD>5</TD><TD>0.7</TD>
        </TR>
        <TR>
          <TD>287.43</TD><TD>-63.85</TD><TD>N 6744</TD><TD>839</TD><TD>6</TD><TD>10.4</TD>
        </TR>
        <TR>
          <TD>023.48</TD><TD>+30.66</TD><TD>N 598</TD><TD>-182</TD><TD>3</TD><TD>0.7</TD>
        </TR>
        </TABLEDATA>
      </DATA>
    </TABLE>
  </RESOURCE>
</VOTABLE>
\end{verbatim}
\ifhtx\End{tabular}
\else
\endgroup
\fi

This simple \elem{VOTABLE} document shows a single \elem{RESOURCE} made of a single \elem{TABLE};
the table is made of 6 columns, each described by a \elem{FIELD}, and has
one additional \elem{PARAM} parameter (the Telescope). The actual rows are
listed in the \elem{DATA} part of the table, here in  XML format 
(introduced by \elem{TABLEDATA}); each cell is marked by the \elem{TD} element, 
and follow the same order as their \elem{FIELD} description:
{\sl RA, Dec, Name, RVel, e\_RVel, R}.

This example also contains a reference to the {\em Space-Time Coordinate} data model
({\em STC}, A. Rots [8]) implicitly used to specify the system of coordinates
used to locate the observed galaxies in the sky:
this is an essential difference from versions of {\em VOTable}
prior to 1.2, which made use of a \elem{COOSYS} element for this specification.
A prescription for relating VOTable metadata to the STC data model is
given in [7].
VOTables which contain appropriate fields are encouraged to declare
their STC metadata in this way if possible.

\subsection{\texorpdfstring{\attr{name}, \attr{ID} and \attr{ref} attributes}
                           {name, ID and ref attributes}}
\label{sec:name}

Most of the elements defined by VOTable may have or have to have {\em names},
like a \elem{RESOURCE}, a \elem{TABLE}, a \elem{PARAM} or a \elem{FIELD}.
The content of the \attr{name} attribute is defined as a {\em token}
XML type,
that is a string of characters where the blanks and spaces are not
meaningful (no leading or trailing spaces, no multiple spaces):
\attrval{name}{NVSS flux(1.4GHz)} represents therefore a 
a valid name.

The \attr{ID} and \attr{ref} attributes are defined as XML types {\em ID}
and {\em IDREF} respectively. This means that the contents of \attr{ID} 
is an {\em identifier} which must be {unique} throughout a VOTable document,
and that the contents of the \attr{ref} attribute represents a reference to 
an identifier which must exist in the VOTable document.
In other terms, if \attrval{ref}{myStar} is found in one element,
there must exist an element in the same document with the
\attrval{ID}{myStar} attribute. The XML standard moreover specifies
that an {\em ID} type is a string beginning with a letter or 
underscore ({\tt{\_}}),
followed by a sequence of Unicode letters, digits, or any of the
punctuation characters {\tt.} (dot), {\tt-} (dash) or {\tt\_} (underscore).
The {\tt:} (colon) is reserved for namespace use and should be avoided.
Therefore \attrval{ID}{1} is {\em not} valid,
but \attrval{ID}{\_1} or \attrval{ID}{ref.1} are both valid.

The {\attr{ID}} attribute 
is therefore required in the elements which {\em have to be referenced},
but the elements having an {\attr{ID}} attribute do not need to be
referenced.
From VOTable 1.2, it is further
recommended to place the \attr{ID} attribute {\em prior to} referencing
it whenever possible.

While the {\attr{ID}} attribute has to be unique in a VOTable document,
the {\attr{name}} attribute need not. It is however recommended,
as a good practice, to assign unique names within a \elem{TABLE} element.
This recommendation means that,
between a \elem{TABLE} and its corresponding closing \elem{\slash TABLE} tag,
{\attr{name}} attributes of \elem{FIELD}, \elem{PARAM} and
optional \elem{GROUP} elements should be all different.

\subsection{\texorpdfstring{\elem{VOTABLE} Element}
                           {VOTABLE Element}}
\label{sec:definitions}

The \elem{VOTABLE} element may contain definitions consisting of
a \elem{DESCRIPTION}, followed by any mixture of parameters and
informative notes eventually structured in {\em groups}.
These elements represent values which are meaningful over all tables
included in a \elem{VOTABLE} document --- definitions specific to
a \elem{RESOURCE} (\Aref{elem:RESOURCE}{section 3.4})
or a \elem{TABLE} (\Aref{elem:TABLE}{section 3.6}) are better placed 
within their most appropriate element.

Note that version 1.0 of VOTable required the usage of a \elem{DEFINITIONS}
element holding the VOTable global definitions --- this
usage is deprecated since version 1.1.

\subsubsection*{Space and Time coordinates}
\label{elem:COOSYS}

An essential difference with the version 1.1 of VOTable concerns 
the way adopted in version 1.2 to describe the {\em coordinate system}:
a dedicated \elem{COOSYS} element was defined in VOTable 1.0, 
which is deprecated in
versions 1.2 and later in  favor of a more generic facility of {\em
referring to external data models}.

The coordinates --- space and time, and eventually the spectral and
redshift parameters --- are described in the {\em STC} model
(A. Rots, see [8]), which specifies the various components
and systems used in Astronomy to locate the events in time and
space with a high accuracy.

Starting with Version 1.2, {\em VOTable} makes use of the 
\elem{GROUP} element (\Arefs{sec:group}{GROUP})  and the \attr{utype} attribute 
(\Arefs{sec:utype}{utype}) to accurately describe 
the coordinate systems used in the data conveyed in a VOTable.
A dedicated note on {\em Referencing STC in VOTable} [7]
describes in more detail how to express the coordinate components.

\subsection{\texorpdfstring{\elem{RESOURCE} Element}
                           {RESOURCE Element}}
\label{sec:resource}
\label{elem:RESOURCE}

A VOTable document contains one or more {\elem{RESOURCE}}
elements, each of these providing a description and the
data values of some logically independent data structure.

Each \elem{RESOURCE} may include the descriptive element {\elem{DESCRIPTION}}, 
followed by a mixture of
{\elem{INFO}}, {\elem{GROUP}} and {\elem{PARAM}} elements;
it may also contain {\elem{LINK}}
elements to provide URL-type pointers that give further information.

The main component of a \elem{RESOURCE} is typically one or more \elem{TABLE}
elements -- in other words a \elem{RESOURCE} is basically a set
of related tables. The \elem{RESOURCE} is recursive (it can contain other
\elem{RESOURCE} elements), which means that the set of tables making up
a \elem{RESOURCE} may become a tree structure.

A \elem{RESOURCE} may have one or both of the \attr{name} or \attr{ID}
attributes (see \Aref{sec:name}{section 3.2}); it may also be qualified by
\attrval{type}{meta}, meaning that the resource is {\em descriptive}
only, i.e. does not contain any actual data: no \elem{DATA} element
should exist in any of its sub-elements. A \elem{RESOURCE} without
this attribute {\em may} however have no \elem{DATA} sub-element.

Finally, the \elem{RESOURCE} element may have a \attr{utype} attribute
to link the element to some external data model 
(introduced in version 1.1, see \Aref{sec:utype}{section 4.6}).

\subsection{\texorpdfstring{\elem{LINK} Element}
                           {LINK Element}}
\label{sec:link}
\label{elem:LINK}

The role of the {\elem{LINK}} element is to provide pointers
to external resources
through a URI. In VOTable, the {\elem{LINK}}
element may be part of a {\elem{RESOURCE}},
{\elem{TABLE}}, \elem{GROUP}, {\elem{FIELD}} or \elem{PARAM} element. 

The linked URI is given by the \attr{href} attribute,
and the nature of the link is indicated by the \attr{content-role} attribute.
The URI should ideally be dereferenceable,
but this is not an absolute requirement,
and appropriate use of the URI depends on the content-role.
This document defines two values for the \attr{content-role} attribute:

\begin{itemize}
\item \attrval{content-role}{doc} indicates documentation.
      Dereferencing the URI should yield a document suitable for
      presentation to the user which describes the LINK's parent element.
      If the URI can produce more than one type, a human-readable response
      must be the default.
      Appropriate behaviour for a client might be to pass the link to a browser
      for presentation.

\item \attrval{content-role}{type} indicates a type-like relationship
      between the URI and the LINK's parent.
      The type is named by the URI string itself,
      while the content retrieved by dereferencing it, if any, is secondary.
      This content-role value would for instance be appropriate
      to mark the LINK's href value as a SKOS concept, e.g.:
      \begin{verbatim}
   <LINK content-role="type"
         href="http://purl.org/astronomy/vocab/PhysicalQuantities/Distance"/>
      \end{verbatim}
\end{itemize}

A \attr{content-role} should be provided for all \elem{LINK} elements,
but if it is absent, a doc-like role may be assumed.
Other values of the \attr{content-role} attribute may be defined
as appropriate outside of this VOTable specification,
for instance by the Semantics Working Group or as part of other
standards that make use of VOTable.

In addition the \elem{LINK} element
may announce the MIME type of the data it references 
with a \attr{content-type} attribute (e.g.\ \attrval{content-type}{image/fits}).
Although this might be overridden by metadata received during the
retrieval operation (e.g.\ the HTTP Content-Type header)
it can serve as a hint to the application about what to expect.

In the Astrores format, from which VOTable is derived, 
there are additional semantics for the {\elem{LINK}}
element; the \elem{href} attribute is used as a template for creating
URLs. This behavior is explained in \Arefx{LINK}{Appendix A.1}, 
and it represents
a possible extension of VOTable.

\subsection{\texorpdfstring{\elem{TABLE} Element}
                           {TABLE Element}}
\label{elem:TABLE}

The \elem{TABLE} element represents the basic data structure in VOTable;
it comprises a description of the table structure (the {\em metadata})
essentially in the form of \elem{PARAM} and \elem{FIELD} elements
(detailed in \Aref{sec:field}{the next section}),
followed by the {\em values} of the described fields in a \elem{DATA}
element (detailed in \Aref{sec:data}{the section below}).

The \elem{TABLE} element is always contained in a \elem{RESOURCE} element:
in other words
any \elem{TABLE} element has a single parent made of the 
\elem{RESOURCE} element
in which the table is embedded. 

The \elem{TABLE} element contains 
a {\elem{DESCRIPTION}} element for descriptive remarks, followed
by a mixed collection of \elem{PARAM}, \elem{FIELD} or \elem{GROUP} elements
which describe a parameter (constant column), a field (column) or a group of
columns respectively. \elem{PARAM} and \elem{FIELD} elements are detailed in 
\Aref{sec:field}{the next section}, and the \elem{GROUP} element
is presented in \Aref{sec:group}{the following section}.

Furthermore the \elem{TABLE} element may contain {\elem{LINK}} elements
that provide URL-type pointers, exactly like the {\elem{LINK}} elements 
existing within a \elem{RESOURCE} element (see \Aref{sec:link}{section 3.5}).

The last element included in a \elem{TABLE} is the optional \elem{DATA} 
element (see \Aref{sec:data}{below}): a table without any
actual data is quite valid, and is typically used to supply a complete
description of an existing resource e.g. for query purposes.

The \elem{TABLE} element may have the naming attributes \attr{name} and/or 
\attr{ID} (see \Aref{sec:name}{name and ID conventions}). A \elem{TABLE}
may also have a \attr{ref} attribute referencing the ID of another
table previously described, which is interpreted as
{\em defining a table having a structure identical to the one referenced}:
this facility avoids a repetition of the definition of tables which
may be present many times in a VOTable document. 
It is recommended that the \attr{ref} attribute
references an {\em empty table} (i.e. a table without a 
\elem{DATA} part), which avoids any ambiguity
about the referencing. 

Finally, the \elem{TABLE} element may have a \attr{utype} and \attr{ucd}
attribute to specify the table semantics, similarly to  the \elem{FIELD} and
\elem{PARAM} elements (see \Aref{elem:FIELD}{section 4.1}).

\section{\texorpdfstring{\elem{FIELD}s and \elem{PARAM}eters}
                        {FIELDs and PARAMeters}}
\label{sec:field}

The atoms of the table structure are represented by \elem{FIELD} and
\elem{PARAM} elements, where \elem{FIELD} represents the description
of an actual table column, while \elem{PARAM} supplies a value
attached to the table, like the \attr{Telescope}
in the example of \Arefs{example1}{section 3.1}. A \elem{PARAM} may be
viewed as a \elem{FIELD} which keeps a {\em constant value} over all
the rows of a table, and the only difference in the set of attributes
of the two elements
is the existence of a \attr{value} attribute in a \elem{PARAM}
which does not exist in a \elem{FIELD}.

The  \elem{FIELD} elements describe the actual columns of the table;
the order in which the \elem{FIELD}s are declared is important,
as this order {\em must} be the same one as the order of the 
columns in \Aref{sec:data}{the data part}.

A {\elem{FIELD}} or \elem{PARAM} element may have several sub-elements, 
including the informational {\elem{DESCRIPTION}}
and {\elem{LINK}} elements (several descriptions and titles
are possible, see \Arefx{sec:addesc}{appendix on additional descriptions}); 
it may also include a {\elem{VALUES}} element
that can express limits and ranges of the values that the
corresponding cell can contain, such as minimum (\elem{MIN}), 
maximum (\elem{MAX}), or
enumeration of possible values (\elem{OPTION}). 

\subsection{Summary of Attributes}
\label{elem:FIELD}
\label{elem:PARAM}
The valid attributes of a \elem{FIELD} or \elem{PARAM} are:

\begin{itemize}
\item   The \attr{name} and/or \attr{ID}. The \attr{ID} attribute is required
        if the field has to be referenced (see 
        \Aref{sec:name}{the generic ID rule}).
        It may help to include the ordinal number of 
        the column in the table in the value of the \attr{ID} attribute
        as e.g. \attrval{ID}{col3} when a single table is involved: 
        the connection to the
        corresponding column would become
        more obvious, especially in the FITS data serialization
        which uses the ordinal column number in the keywords containing
        the metadata related to that column. 

\item   The \attr{datatype}, which expresses the nature of the data 
        that is described as one of the permitted primitives 
        (see \Tref{primitives}{the table above} and their exact meaning 
        in \Aref{sec:datatypes}{section 6}).
        This attribute determines
        how data are read and stored internally;
        it is {\em required}. 

\item   The \attr{arraysize} attribute exists when 
        the corresponding table cell contains more than one of the specified
        datatype, as explained in \Aref{sec:dim}{section 2.2}.
        Note that strings are not a primitive type,
        and have to be described as an array of characters.

\item   {\fg{black}}The \attr{width} and \attr{precision} attributes define the 
        numerical accuracy associated with the data 
        (see \Aref{sec:form}{below}).

\item	The \attr{xtype} attribute, added in VOTable 1.2, specifies an
	{\em extended} (or {\em external}) datatype. It is meant
	to give details about the column contents beyond the 
	primitive \attr{datatype}\revision{2}{, like timestamps.}
	{; a typical example is 
	\attrval{type}{iso8601} for timestamps.}

\item   The \attr{unit} attribute specifies the units in which
        the values of the corresponding column are expressed
        (see \Aref{sec:unit}{below})

\item   The \attr{ucd} attribute supplies a standardized classification
        of the physical quantity expressed in the column
        (see \Aref{sec:ucd}{below}).

\item   The \attr{utype} attribute, introduced in VOTable 1.1, is meant
        to express the role of the column in the context of an external
        data model (see \Aref{sec:utype}{below}); it is used in
        the example  \Aref{example1}{above} to specify {\em which 
        coordinate component} a field represents, in connection with
        the \attr{ref} attribute.

\item   The \attr{ref} attribute is used to quote another element of
        the document in the definition of a \elem{FIELD} or \elem{PARAM}. 
        It is used in the example of \Aref{example1}{the example} 
        to indicate the coordinate system in which the coordinates 
        are expressed
        (reference to the \elem{GROUP} element which specifies the
        coordinate frame).

\item   The \attr{type} attribute is {\em not} part of this standard,
        but is reserved for future extensions (see 
        \Arefx{LINK}{Link substitution},
        \Arefx{query}{Query Extension} and 
        \Arefx{location}{fields as pointers}).

\end{itemize}

In addition, in the \elem{PARAM} element only:
\begin{itemize}
\item   the \attr{value} attribute which explicits the \elem{PARAM}eter's
        value; \attr{value} is a required attribute of the \elem{PARAM}
        element.
\end{itemize}

\subsection{Numerical Accuracy}
\label{sec:form}

The VOTable format is meant for transferring, storing, and
processing tabular data, and is not intended for presentation
purposes: therefore (in contrast to Astrores) we generally avoid
giving rules on presentation, such as formatting. 
Inevitably however at least some of the data will be presented --
either as actual tables,  or in forms or graphs, etc.
Two attributes were retained for this purpose:

\begin{itemize}
\item   The {\attr{width}} attribute 
        is meant to indicate to the application
        the number of characters to be used for input
        or output of the quantity. 

\item   The \attr{precision} attribute is meant to express the
        number of significant digits, either as a number of
        decimal places (e.g. \attrval{precision}{F2} or equivalently
        \attrval{precision}{2} to express 2 significant figures
        after the decimal point), or as a number of significant figures
        (e.g. \attrval{precision}{E5} indicates a relative precision
        of $10^{-5}$).
\end{itemize}

The existence and presentation of the special {\em null} value of 
a field (when the actual value of the field is unknown) is
another aspect of the numerical accuracy, which is part of the
\elem{VALUES} sub-element (see \Aref{sec:values}{below}).

\subsection{\texorpdfstring{Extended Datatype \attr{xtype}}
                           {Extended Datatype xtype}} \label{sec:xtype}

The \attr{xtype} attribute expands the basic
datatype primitives (in \Tref{primitives}{table of primitives})
representing the storage units which are valid in any of the 
VOTable serializations,
and corresponds therefore exactly to the {\em FITS} definitions.
It fills the gap between the datatypes
known by FITS and those required to express queries 
(Astronomical Data Query Language or ADQL, see [12])
and their results in tabular form (Table Access Protocol or TAP, 
see [11]).

The \attr{xtype} attribute is the way to 
specify that a parameter represents a {\em  timestamp} 
(an instant in an absolute time frame), materialized by a 
UTC date/time string following the ISO-8601 standard 
({\tt YYYY-MM-DDThh:mm:ss} eventually followed by a decimal point 
and fractions of seconds);
parameters required to specify a spatial position may also have an associated
\attr{xtype}.

The actual values of the \attr{xtype} attribute are not defined
in this VOTable specification; it is expected however that 
common conventions will be adopted by the various components
of the Virtual Observatory, in a way similar to the adoption of the
Unified Content Descriptor (\Arefs{example1}{section 4.5})

\subsection{Units}
\label{sec:unit}

The quantities in a column of the table may be expressed in
some physical unit,
which is specified by the {\attr{unit}}
attribute of the {\elem{FIELD}}.
The  syntax of the {\em unit} string is defined in reference [3];
it is basically written as a string without blanks or spaces,
where the symbols {\bf.} or {\bf*} indicate a multiplication,
{\bf/} stands for the division, and no special symbol is required
for a power.
Examples are \attrval{unit}{m2} for m$^2$,
\attrval{unit}{cm-2.s-1.keV-1} for cm$^{-2}$s$^{-1}$keV$^{-1}$,
or \attrval{unit}{erg/s} for erg\,s$^{-1}$.
The references [3] provide also the list of the valid symbols,
which is essentially restricted to the {\em Syst\`eme International}
(SI) conventions, plus a few astronomical extensions concerning
units used for time, angle, distance and energy measurements.

\subsection{Unified Content Descriptors}
\label{sec:ucd}

The Unified Content Descriptors (UCD) can be viewed as a 
hierarchical glossary of the scientific meanings of the data 
contained in the astronomical tables.
Two versions of UCDs have been developed:
the initial version (UCD1) created at CDS, which uses
atomic words separated by underscores (e.g. {\tt POS\_EQ\_RA\_MAIN});
and a more flexible one, UCD1+ [4],
developed in the frame of the IVOA Semantics Working Group, which uses
a reduced vocabulary of dot-separated atoms which can be
combined with semi-colons (e.g. {\tt pos.eq.ra;meta.main}).
UCD1+ usage is recommended, but applications using the older
vocabulary are still acceptable in this version of VOTable.

\noindent A few typical examples of UCD1+ definitions
 are:

\begin{tabular}{ll}
{\literalvalue{phot.mag;em.opt.B}}        &  Blue magnitude \\
{\literalvalue{src.orbital.eccentricity}} &  Orbital eccentricity \\
{\literalvalue{time.period;stat.median}}  &  Median Value of the Period \\
{\literalvalue{instr.det.qe}}             &  Detector's Quantum Efficiency \\
\end{tabular}

\subsection{\texorpdfstring{The \attr{utype} Attribute}
                           {The utype Attribute}}
\label{sec:utype}
In many contexts, it is important to specify that \elem{FIELD}s or 
\elem{PARAM}eters convey the values defined in an external {\em data
model}. For instance, it can be fundamental for an application to
be aware that a given \elem{FIELD} expresses { the} surface brightness
measured with a specific filter and within a 12x6arcsec elliptical aperture.
None of the other \attr{name}, \attr{ID}
or \attr{ucd} attributes can fill this role, and 
the \attr{utype} (usage-specific or {\em unique} type) attribute was 
introduced in VOTable 1.1 to fill this gap. 
By extension, most elements may refer to some external data model,
and the \attr{utype} attribute is also legal in \elem{RESOURCE},
\elem{TABLE} and \elem{GROUP} elements.

In order to avoid name collisions, the data model identification
should be introduced following the XML namespace conventions,
as \attrval{utype}{{\rm\em datamodel\_identifier:role\_identifier}}.
The mapping of \literalvalue{datamodel\_identifier} to an xml-type attribute
is recommended, by means of the {\tt xmlns} convention
which specifies the URI of the data model cited, as done
in the example of \Arefs{example1}{section 3.1}.

The {\tt utype} attribute is especially useful to  specify
the {\em spatial and temporal coordinates} present  in the table
when it contains astronomical events: these parameters 
are  essential to most applications which process multi-wavelength data.
Within the IVOA, the spatial and temporal
frames are described in the {\bf STC} data model (see Rots [8]),
and it is expected that this {\em STC}-referencing replaces
the usage of the \elem{COOSYS} defined in the version 1.0 of VOTable.

The example given above (see \Arefs{example1}{section 3.1})
gives an illustration of the recommended way of linking
a VOTable document to the STC model. Other examples and
details are presented in the dedicated note
{\em``Referencing STC in VOTable''} [7].

\subsection{\texorpdfstring{\elem{VALUES} Element}
                           {VALUES Element}}
\label{sec:values}

The {\elem{VALUES}} element of the {\elem{FIELD}}
is designed to hold subsidiary information about the {\em domain} of the
data. For instance, in the example (\Aref{example1}{section 3.1})
we could rewrite the RA field definition as:

\ifhtx\Beg{tabular}{\bg{LightCyan} CELLPADDING=5}{||l||}
\else\begingroup\small\fi
\begin{verbatim}
      <FIELD name="RA" ID="col1" ucd="pos.eq.ra;meta.main"
             datatype="float" width="6" precision="2" unit="deg">
        <VALUES ID="RAdomain">
          <MIN value="0"/>
          <MAX value="360" inclusive="no"/>
        </VALUES>
      </FIELD>
\end{verbatim}
\ifhtx\End{tabular}\else\endgroup\fi

\noindent The scope of the domain described by the \elem{VALUES} element
(and by its \elem{MIN}, \elem{MAX} and \elem{OPTION} sub-elements)
can be qualified by \attrval{type}{actual}, if it is valid only for
the data enclosed in the parent \elem{TABLE}; the default \attrval{type}{legal} 
qualification specifies the generic domain of valid values, as in the
{\em RAdomain} in the example above where the interval $[0,360[$ is specified.

\label{elem:VALUES}
\label{elem:MIN}
\label{elem:MAX}
\label{elem:OPTION}
The \elem{VALUES} element may contain {\elem{MIN}} and {\elem{MAX}} elements, 
and it may contain {\elem{OPTION}} elements; 
the latter may itself contain more {\elem{OPTION}}
elements, so that a hierarchy of keyword-values pairs can be
associated with each field. 
Note that only a single pair \elem{MIN} / \elem{MAX} is possible, 
whereas many \elem{OPTION} elements may be used to qualify the domain
described by the \elem{VALUES} element. 
The domain may therefore be defined as a single interval, or as a set
of individual values. Although the schema does not forbid all three
\elem{MIN}, \elem{MAX} and \elem{OPTION} sub-elements simultanesouly, 
such usage is considered as bad practice and is discouraged.

All three \elem{MIN}, \elem{MAX} and \elem{OPTION} sub-elements 
store their value corresponding to the minimum, maximum, or ``special value''
in a \attr{value} attribute. \elem{MIN} and \elem{MAX} elements
can have an \attr{inclusive} attribute to specify whether the \attr{value}
quoted belongs to the domain or not, and the  \elem{OPTION} element
can have a \attr{name} attribute to describe the ``special'' quoted 
\attr{value}.

The \elem{VALUES} element may also have a \attr{null} attribute 
to define a non-standard value that is used to specify 
{\em``non-existent data''} -- for example \attrval{null}{-32768}.
When this value is found in the corresponding data, it is assumed that no data
exists for that table cell; the parser may also choose to use this
when unparsable data is found, and the null value will be substituted
instead.
The value of the \attr{null} attribute must follow the same rules
as the TABLEDATA serialization for the appropriate datatype
described in \Aref{sec:datatypes}{section 6},
and may never contain an array value.
This mechanism is only intended for use with integer types;
it should not be used for floating point types, which can use NaN instead.

This mechanism for representing null values is required for integer
columns in the \elem{BINARY} serialization.
Since VOTable 1.3 however other mechanisms are available for representing
null values in the \elem{TABLEDATA} and \elem{BINARY2} serializations.
Representation of nulls using the \elem{VALUES} element and otherwise
is discussed further in \Aref{sec:NULL}{section 5.5}.

Finally the \attr{ref} attribute of a \elem{VALUES} element
can be used to avoid a repetition of the domain definition,
by referring to a previously defined \elem{VALUES} element
having the referenced \attr{ID} attribute. 
When specified, the  \attr{ref} attribute completely defines
the domain without any other element or attribute, e.g.
\elemdef{VALUES}{ \attrval{ref}{RAdomain}\slash}.

\subsection{\texorpdfstring{\elem{INFO} Element}
                           {INFO Element}}
\label{elem:INFO}
The \elem{INFO} element is a {\elem{PARAM}} element restricted 
to be of type {\em string}
(i.e. \attrval{datatype}{char} and \attrval{arraysize}{*} are {\em implied}).
It {\em must} also have a \attr{name} attribute,
and {\em may} have the other attributes allowed in a \elem{PARAM}:
\attr{ID}, \attr{ref}, \attr{unit}, \attr{ucd} and \attr{utype}.
But unlike \elem{PARAM}, \elem{INFO} does not accept sub-elements:
only text is acceptable in \elem{INFO}'s body. This limitation ensures full 
compatibility with the previous versions of VOTable.

\elem{INFO} is meant to convey informative details about the
generation of the {VOTABLE} document.
It may be present
at the beginning or end of \elem{VOTABLE} or \elem{RESOURCE} elements,
or at the end of a \elem{TABLE}. Typical uses of \elem{INFO}
include error reports, or explanations about choices made by the
data processing system which generates the VOTable document.

\subsection{\texorpdfstring{\elem{GROUP}ing \elem{FIELD}s and \elem{PARAM}eters}
                           {GROUPing FIELDs and PARAMeters}}
\label{sec:group}
\label{elem:GROUP}
\label{elem:FIELDref}
\label{elem:PARAMref}

The \elem{GROUP} element is used 
to group together a set of \elem{FIELD}s and \elem{PARAM}s
which are logically connected, like a value and its error. 
The \elem{FIELD}s are always defined {\em outside} any group,
and the \elem{GROUP} designates its member fields via  
\elem{FIELDref} elements.

A simple example of a group made of the velocity and its error, 
based on the example of \Aref{example1}{section 3.1}, 
can be the following:

\ifhtx\Beg{tabular}{\bg{LightCyan} CELLPADDING=5}{||l||}
\fi
\begin{verbatim}
    <GROUP name="Velocity">
      <DESCRIPTION>Velocity and its error</DESCRIPTION>
      <FIELDref ref="col4"/>
      <FIELDref ref="col5"/>
    </GROUP>
\end{verbatim}\ifhtx\End{tabular}\fi

The \elem{GROUP} element can have the \attr{name}, \attr{ID}, \attr{ucd},
\attr{utype} and \attr{ref} attributes.
It can include a \elem{DESCRIPTION}, and any mixture of 
\elem{FIELDref}erences,
\elem{PARAM}eters, \elem{PARAMref}erences
and other \elem{GROUP}s. \elem{PARAMref} is a {\em logical} definition
of a parameter that refers to a \elem{PARAM}
element defined elsewhere in the parent \elem{TABLE} or \elem{RESOURCE};
similarly the \elem{FIELDref} element defined by referring
to a \elem{FIELD} element defined elsewhere in the parent \elem{TABLE}.
The recursivity of the \elem{GROUP} element enables a definition of
arbitrarily complex structures.

The possibility of adding \elem{PARAM}eters in groups also introduces
a possibility of associating parameter(s) to  accurately  describe
the context of the data stored in the table. 
For instance,
it is possible to associate the actual frequency of a radio survey with
a table of flux measurements using
the following declaration:

\ifhtx\Beg{tabular}{\bg{LightCyan} CELLPADDING=5}{||l||}\fi
\begin{verbatim}
    <FIELD name="Flux" ID="col4" ucd="phot.flux;em.radio.200-400MHz" 
           datatype="float" width="6" precision="1" unit="mJy"/>
    <FIELD name="e_Flux" ID="col5" datatype="float" width="4" precision="1"
           ucd="stat.error;phot.flux;em.radio.200-400MHz" unit="mJy"/>
    <GROUP name="Flux" ucd="phot.flux;em.radio.200-400MHz">
      <DESCRIPTION>Flux measured at 352MHz</DESCRIPTION>
      <PARAM name="Freq" ucd="em.freq" unit="MHz" datatype="float" 
             value="352"/>
      <FIELDref ref="col4"/>
      <FIELDref ref="col5"/>
    </GROUP>
\end{verbatim}\ifhtx\End{tabular}\fi

\par
Similarly, \elem{GROUP} can be used to associate several parameters
to one or several \elem{FIELD}s. For example, a filter may be
characterized by the central wavelength and the FWHM of its transmission
curve, or several parameters of an instrument setup may be described.

\subsection{The Relational Context}
\label{sec:relation}

With a simple naming convention, 
the \elem{GROUP} element may also specify some
properties of the tables included in a VOTable document
when a \elem{TABLE} is viewed as a {\em relation} (part of a
a relational database):

\begin{itemize}
\item   A \elem{GROUP} element having the \attrval{name}{primaryKey} 
        attribute defines the {\em primary key} of the relation
        by enumerating the ordered list of \elem{FIELDref}s that 
        make up the {\em primary key} of the table;
\item   A \elem{GROUP} element having the \attrval{name}{foreignKey}
        attribute, with a \attrval{ref}{{\rm\em table\_reference}}
        reference of the table having the associated primary ley,
        similarly enumerates the \elem{FIELDref}s of the
        {\em foreign key};
\item   A \elem{GROUP} element having the \attrval{name}{order}
        attribute may specify how the data are ordered.
\end{itemize}

\noindent Similar conventions could be added for 
the existence of indexes, unique values, etc.

\section{Data Content}
\label{sec:data}

While the bulk of the metadata of a VOTable document is in the
{\elem{FIELD}} elements, the data content of the table is 
in a single {\elem{DATA}} element. 
The data is organized in ``reading" order, so that
the content of each row appears in the same order as the order of the
{\elem{FIELD}} definitions.

\label{Image1}
\ifhtx\tag{IMG SRC="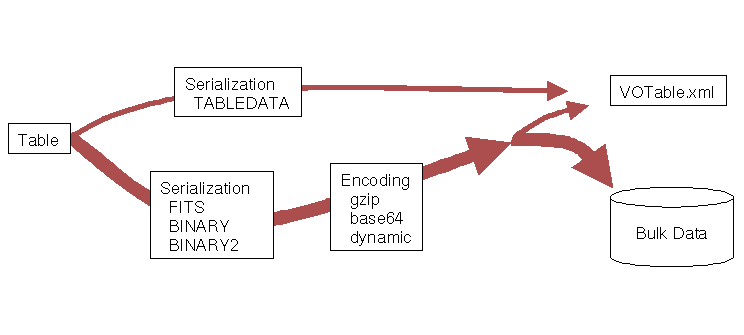" NAME="Image1" ALIGN=BOTTOM BORDER=0}\br
\else
\begin{figure}[hbt]
\includegraphics[width=\textwidth]{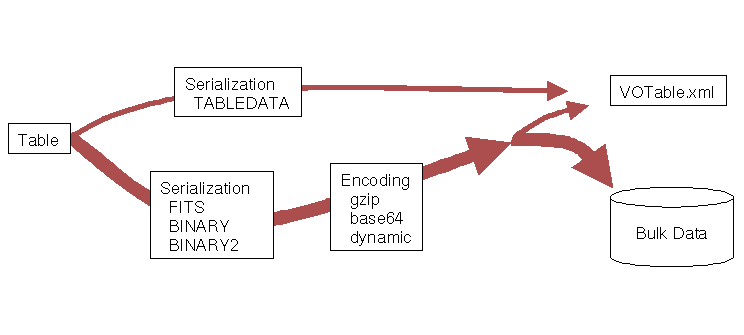}
\caption{\label{fig:serialization}Data serialization}
\end{figure}
\fi

Each \elem{DATA} part of the VOTable document can be viewed as
a stream coming out of a pipeline.
The abstract table is first serialized by one of several
methods, then it may be encoded for compression or other reasons. The
result may be embedded in the XML file ({\it local} data), or it may
be {\it remote} data.

\Fref{fig:serialization}{The figure}
shows how the abstract table is rendered into the
VOTable document. First the data is {\it serialized}, either
as XML, a FITS binary table, or the VOTable
Binary format. This data stream may then be {\it encoded},
perhaps for compression or to convert binary to text. Finally, the
data stream may be put in a remote file with a URL-type pointer in
the VOTable document; or the table data may be embedded in the
VOTable.

The serialization elements and their attributes are
described in the next sections.

\subsection{\texorpdfstring{\elem{TABLEDATA} Serialization}
                           {TABLEDATA Serialization}}
\label{sec:TABLEDATA}
\label{elem:TD}
\label{elem:TR}

The \elem{TABLEDATA} element is a way to build the table in pure XML,
and has the advantage that XML tools can manipulate and present
the table data directly.
The \elem{TABLEDATA} element  contains {\elem{TR}}
elements, which in turn contain {\elem{TD}}
elements --- i.e. the same conventions as in HTML.
The number of {\elem{TD}} elements in each {\elem{TR}} element
must be equal to the number of {\elem{FIELD}} elements declaring the table.
An example is contained in \Aref{example1}{section 3.1},
surrounded by the \elemdef{TABLEDATA}{} and \elemdef{\slash TABLEDATA}{}
delimiters.

Each item in the {\elem{TD}} tag 
contains a value which must be compatible with 
the {\attr{datatype}} attribute of the corresponding {\elem{FIELD}} definition.
If the value is the same as the {\attr{null}} value for that field
(see \Aref{sec:values}{section 4.7})
then the item is assumed to contain no data.
Valid representations of values in a cell, depending on their
\attr{datatype}, are detailed in \Aref{sec:datatypes}{the complete
description of datatypes}.
If the {\elem{TD}} element is empty ({\verb|<TD/>| or \verb|<TD></TD>|)
the cell is considered to contain no data, i.e.\ to be null.

If a cell contains an array of numbers or a complex number, 
it should be encoded as multiple numbers separated by
whitespace. However in the case of character and Unicode strings
(declared in the corresponding \elem{FIELD} as an array of {\em char} 
or {\em unicodeChar} datatype), no
separator should exist. Here is an example of a two-row table 
that  has arrays in the table cells:

\ifhtx\Beg{tabular}{\bg{LightCyan} CELLPADDING=5}{||l||}
\else\begingroup\small\fi
\label{example2}
\begin{verbatim}
<TABLE>
  <FIELD name="aString" datatype="char" arraysize="10"/>
  <FIELD name="aShort"  datatype="short"/>
  <FIELD name="varInts" datatype="int"  arraysize="*"/>
  <FIELD name="Floats"  datatype="float"arraysize="3"/>
  <DATA><TABLEDATA>
    <TR> <TD>Apple</TD>  <TD/>       <TD>1 2 4 8 16</TD> <TD>1.62 4.56 3.44</TD> </TR>
    <TR> <TD>Orange</TD> <TD>15</TD> <TD>23 -11 9</TD>   <TD>2.33 4.66 9.53</TD> </TR>
  </TABLEDATA></DATA>
</TABLE>
\end{verbatim}
\ifhtx\End{tabular}\else\endgroup\fi

The first entry is a fixed-length array of 10 characters; since
the value being presented ({\tt Apple}) has 5 characters, this
is padded with trailing blanks. The second cell is a short integer
but has a null value, as indicated by the empty \elem{TD} element.
The third cell contains a variable-length array of integers.
The last cell contains a fixed-length array of three floats.

A special notice should be mentioned about the significance of 
{\em white space} in a table cell (the term {\em white space} 
designates the characters {\em space} [{\tt{x20}}], {\em tab} [{\tt{x09}}],
{\em newline} [{\tt{x0a}}], {\em carriage-return} [{\tt{x0d}}]):
while for numeric data types
the amount of white spaces does not matter (the elements
of an array of numbers may for instance be written on several lines),
the white space is significant for \literalvalue{char} or
\literalvalue{unicodeChar} datatypes, and for instance
{\fg{DarkPurple}\verb+<TD>Apple</TD>+} and 
{\fg{DarkPurple}\verb+<TD> Apple</TD>+} are {\em not} identical.

\subsection{\texorpdfstring{\elem{FITS} Serialization}
                           {FITS Serialization}}
\label{sec:FITS}
\label{elem:FITS}

The FITS format for binary tables [2] is in widespread use in astronomy,
and its structure has had a major influence on the VOTable specification.
Metadata is stored in a header section, followed by the data. The
metadata is essentially equivalent to the metadata of the VOTable
format. One important difference is that VOTable does not require
specification of the number of rows in the table, an important
advantage if the table is being created dynamically from a stream.

The VOTable specification does not define the behavior of parsers
with respect to this doubling of the metadata. A parser may ignore
the FITS metadata, or it may compare it with the VOTable metadata for
consistency, or other possibilities.

The following code shows a fragment that might have been created
by a FITS-to-VOTable converter. Each FITS keyword has been converted
to a \elem{PARAM}, and the data itself is remotely stored and gzipped at an
FTP site:

\begin{plain}\small
\elemdef{RESOURCE}{}\\
\hspace*{0.5em}\elemdef{PARAM}{ \attrval{name}{EPOCH} \attrval{datatype}{float} 
        \attrval{value}{1999.987}}\\
        \hspace*{1em}\elemdef{DESCRIPTION}{}
	Original Epoch of the coordinates\elemdef{\slash DESCRIPTION}{}\\
\hspace*{0.5em}\elemdef{\slash PARAM}{} \\
\hspace*{0.5em}\elemdef{PARAM}{ \attrval{name}{TELESCOP} \attrval{datatype}{char} 
   \attrval{arraysize}{*} \attrval{value}{VTel} \slash}\\
\hspace*{0.5em}\elemdef{INFO}{ \attrval{name}{HISTORY}}\\
  \hspace*{1em}
  The very first Virtual Telescope observation made in 2002\\
\hspace*{0.5em}\elemdef{\slash INFO}{} \\
\hspace*{0.5em}\elemdef{TABLE}{} \\
\hspace*{1.0em}\elemdef{FIELD}{{\rm\em\quad(insert field metadata here)} \slash}\\
\hspace*{1.0em}\elemdef{DATA}{}\elemdef{FITS}{ \attrval{extnum}{2}}\\
\hspace*{1.5em}\elemdef{STREAM}{ \attrval{encoding}{gzip} 
           \attrval{href}{ftp://archive.cacr.caltech.edu/myfile.fit.gz}\slash}\\
\hspace*{1.0em}\elemdef{\slash FITS}{}\elemdef{\slash DATA}{} \\
\hspace*{0.5em}\elemdef{\slash TABLE}{}\\
\elemdef{\slash RESOURCE}{}
\end{plain}

The FITS file may contain many data objects (known as extensions, 
numbered from 1 up, the main header being numbered 0), and the
\attr{extnum} attribute allows the VOTable to point to one of
these.

\subsection{\texorpdfstring{\elem{BINARY} Serialization}
                           {BINARY Serialization}}
\label{sec:BIN}

The binary format is intended to be easy to read by parsers, so
that additional libraries are not required. It is just a sequence of
bytes with the length of each sequence corresponding to the {\attr{datatype}}
and {\attr{arraysize}} attributes of the {\elem{FIELD}}
elements in the metadata. The binary format consists of a sequence of
records with no header bytes, no alignment considerations, and no block sizes.
The order of the bytes in multi-byte primitives (e.g. integers,
floating-point numbers) is Most Significant Byte first, i.e.,
it follows the FITS convention.

Table cells may contain arrays of primitive types, each of which
may be of fixed or variable length. In the former case, the number of
bytes is the same for each instance of the item, as specified by the
{\attr{arraysize}}
attribute of the {\elem{FIELD}}.
If all the fields have a fixed {\attr{arraysize}},
then each record of the binary format has the same length
(the sum of {\attr{arraysize}}
times the length in bytes of the corresponding {\attr{datatype}}).

Variable-length arrays of primitives are preceded by a 4-byte integer
containing the number of items of the array.
The parser can then compute the number of bytes taken
by the variable-length array by multiplying the size and number 
of the primitives.

The way the stream of bytes is arranged for the data of the 
example in \Aref{example2}{section 5.1} is illustrated in 
\Fref{fig:bin}{Figure 2}.
In this case the second column must be declared like this:
\begin{verbatim}
  <FIELD name="aShort" datatype="short">
    <VALUES null="99"/>
  </FIELD>
\end{verbatim}
to indicate a magic value representing nulls, since no equivalent of the
empty \elem{TD} element is available for the BINARY serialization
(see \Aref{sec:NULL}{section 5.5}).

\label{Image2}
\ifhtx\begin{tabular}{c}
\tag{IMG SRC="binary.png" NAME="Image2" ALIGN=LEFT BORDER=0}\end{tabular}
\else\begin{center}
\begin{figure}[htb]
\includegraphics[width=\textwidth]{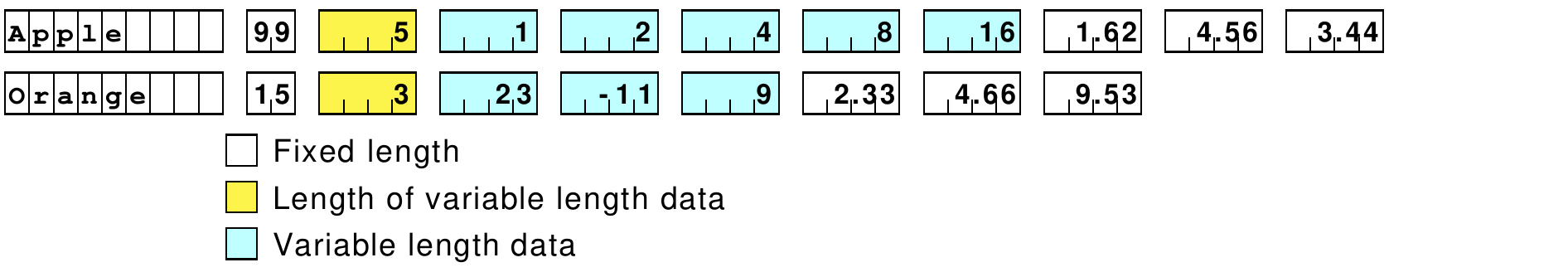}
\caption{\label{fig:bin}Data Storage in BINARY mode}
\end{figure}\end{center}
\fi

The BINARY serialization has been available in all versions of VOTable.
From VOTable 1.3 however, the alternative BINARY2 serialization is
an alternative, providing more straightforward null-flagging capabilities.
In VOTable 1.3 BINARY remains a legal serialization, but for most
purposes VOTable producers are advised to use BINARY2 instead.

\subsection{\texorpdfstring{\elem{BINARY2} Serialization}
                           {BINARY2 Serialization}}
\label{sec:BIN2}

The BINARY2 format, introduced at VOTable 1.3, is the same as BINARY,
but with null entries flagged explicitly rather than identified
by their values.
The byte stream contains one additional bit for each table cell
indicating whether that cell's value is to be considered null or not.

The byte content for each row consists of
zero or more bytes containing a null value flag for each cell in the row,
followed by the bytes for the BINARY serialization as described in the
previous subsection.
The null flags are stored as exactly one bit per table column, and the number
of flag bytes is the smallest required for this purpose;
the number of flag bytes per row for an $N$-column table will therefore
be the integer part of  $(N+7)/8$.
The most significant bit of the first flag byte corresponds to the
first column,
the second most significant bit of the first flag byte to the second column,
the most significant bit of the second flag byte to the eighth column,
and so on.  A set (1) bit indicates that the corresponding cell is null,
and an unset (0) bit indicates that its value is not null.
Unused bits will be at the less-significant end of the final flag byte,
and shall be unset (0).

It is recommended, but not required, that a cell value flagged as null
is filled with with the NaN value for floating point or complex datatypes,
and zero-valued bytes for other datatypes.
It is particularly recommended that a variable length array cell value
flagged as null is represented as 4 zero-valued bytes, indicating
a zero-length value.

The way the stream of bytes is arranged for the data of the
example in \Aref{example2}{section 5.1} is illustrated in
\Fref{fig:bin2}{Figure 3}.

\label{Image3}
\ifhtx\begin{tabular}{c}
\tag{IMG SRC="binary2.png" NAME="Image3" ALIGN=LEFT BORDER=0}\end{tabular}
\else\begin{center}
\begin{figure}[htb]
\includegraphics[width=\textwidth]{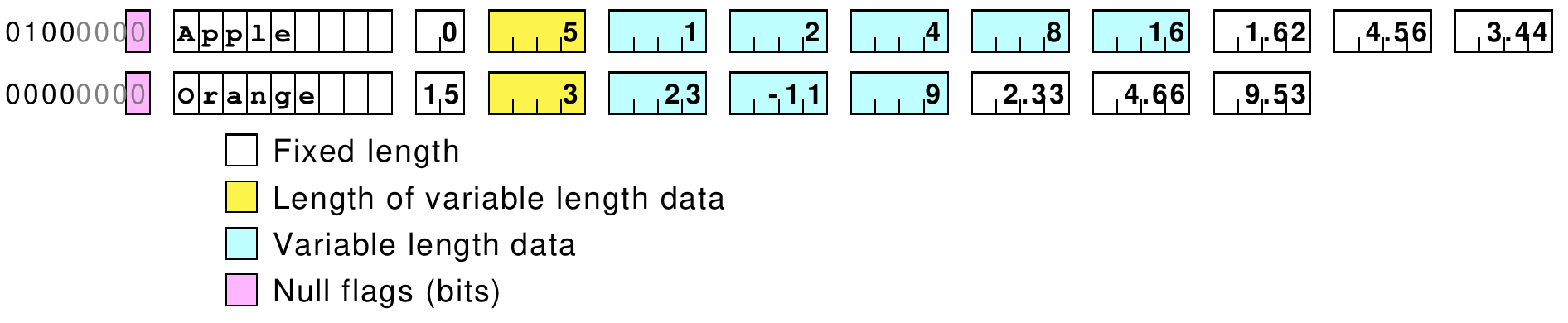}
\caption{\label{fig:bin2}Data Storage in BINARY2 mode}
\end{figure}\end{center}
\fi

\subsection{Null values}
\label{sec:NULL}

VOTable provides two approaches to representing null values in data.

The first approach makes use of the \elem{VALUES} element's \attr{null}
attribute to indicate that whenever a particular ``magic'' value is
encountered in a column's data, that entry should be considered as null.
This magic value must represent a legal scalar value for the column's
datatype, for instance in the case of {\attrval{datatype}{unsignedByte}}
it must be an integer in the range 0--255.
This approach, inherited from FITS, works in the same way for all
of the defined VOTable serializations.
However it can present difficulties when generating VOTables,
since the magic value must be distinct from all actual data values
in the column, and must be chosen before the column data has
been written, since the \elem{FIELD} element precedes the \elem{DATA}.

The second approach, introduced at VOTable 1.3,
is to mark null values using some mechanism external
to the data itself, and it works differently for the different serializations.
In the \elem{TABLEDATA} serialization
an empty \elem{TD} element signals a null value, and
in the \elem{BINARY2} serialization
a separate null-ness flag is provided for each cell.
The \elem{BINARY} and \elem{FITS} serializations do not support this approach
at all.
It should be noted that when using this approach, unlike with magic values,
the different serializations do not have identical capabilities for
representing data, so that lossless round-tripping between serializations
is not always possible.

Some other subtleties concerning null values should also be mentioned:
\begin{itemize}
\item The only way to mark as null individual elements of an array-valued cell
      is by use of the magic value mechanism, which operates on a
      per-element basis.
      Although the magic value approach can mark individual elements
      of an array as null, it cannot mark a whole multi-element array as null.
\item In TABLEDATA array-valued columns, a null value and a zero-length
      array are not distinguished.
      Since strings are represented as arrays of characters, this also
      means that empty and null strings are not distinguished.
\item In either approach, floating point values not formally marked as nulls
      may take the value NaN (not-a-number),
      represented by the string ``NaN'' or by a suitable IEEE bit pattern
      as appropriate.
      This option is suitable for scalar, complex, and array-valued columns.
      For most purposes, the distinction between NaN and null is not
      significant, and VOTable implementations are not required to distinguish
      these cases.  However, the BINARY2 encoding does provide the option
      to represent them differently for specialised applications where
      that is desirable.
\item The magic value mechanism, as in FITS, is only intended for integer
      values.  Historically it has not been explicitly forbidden for
      floating point values, but such use is strongly deprecated in favour
      of the use of NaN.
\item Combining the two approaches is not encouraged,
      and use of the \elem{VALUES} \attr{null} attribute is deprecated
      where it can be avoided (marking null cells in TABLEDATA and BINARY2
      serializations in VOTable 1.3).
      However, if it is present, the \elem{VALUES} \attr{null} attribute
      must always be respected.
\item The {\em boolean} datatype has its own arrangements for representing
      null which do not require use of either of the special approaches above.
\end{itemize}

\subsection{Data Encoding}
\label{elem:STREAM}

As a result of the serialization, the table has been converted to
a byte stream, either text or binary. If the {\elem{TABLEDATA}}
serialization is used, then the table is represented as XML tags 
directly  embedded in the document,
and conventional tools can be used to encode the entire XML document.
However, VOTable also provides limited encoding of its own. 
A VOTable document may point to a remote data resource that is compressed; 
rather than decompressing before sending on the wire, it can be dynamically
decoded by the VOTable reader. We might also use the encoding facilities to 
convert a binary file to text (through base64 encoding), so that binary 
data can be used in the XML document.

In this version (1.3) of VOTable, it is not possible to encode
individual columns of the table: the whole table must be encoded in
the same way. However, the possibility of encoding selected table cells
is  being examined for future versions of VOTable
(see \Arefx{sec:b64}{appendix below}).

In order to use an encoding of the data, it must be enclosed in a
{\elem{STREAM}}
element, whose attributes define the nature of the encoding. The
{\attr{encoding}}
attribute is a string that should indicate to the parser how to undo
the encoding that has been applied. Parsers should understand and
interpret at least the following values:
\begin{itemize}
        \item {\attrval{encoding}{gzip}} [RFC1952]
        implies that the data following has been compressed with the {\em gzip}
        filter, so that {\em gunzip} or similar should be applied.
        \item {\attrval{encoding}{base64}} [RFC2045]
        implies that the {\em base64} filter has been applied, to convert binary
        to text.
        \item {\attrval{encoding}{dynamic}}
        implies that the data is in a remote resource (see below), and the
        encoding will be delivered with the header of the data.
        This occurs with the http protocol, where the MIME header indicates 
        the type of encoding that has been used.
\end{itemize}

\noindent The default value of the encoding attribute is the null string, 
meaning that no encoding has been
applied. In future releases, we might allow more complex strings in
the encoding attribute, allowing combinations of encoding filters and
a way for the parser to find the software needed for the decoding.

Note that for inline streamed data
(a \elem{STREAM} with no \attr{href} attribute)
it is effectively required to use \attrval{encoding}{base64},
since of the available options only base64 will ensure that
binary data is encoded as legal XML content.

\subsection{Remote Data}

If the encoding of the data produces text, or if the serialization
is naturally text-based, then it can be directly embedded into the
XML document:
\begin{plain}
\hspace*{0.5em}\elemdef{DATA}{}\elemdef{BINARY}{}\\
\hspace*{1em}\elemdef{STREAM}{ \attrval{encoding}{base64}}\\
\hspace*{1.5em}{\tt
\verb+AAAAAj/yVZiDGSSUwFZ6ypR4yGkADwAcQV0euAAIAAJBmMzNwZWZmkGle4tBR3jVQT9ocwAA+
}\\
\hspace*{1.5em} $\cdots\cdots\cdots\cdots\cdots\cdots\cdots\cdots$\\
\hspace*{1em}\elemdef{\slash STREAM}{}\\
\hspace*{0.5em}\elemdef{\slash BINARY}{}\elemdef{\slash DATA}{}
\end{plain}

However, if the data stream is very large, it may be preferable to keep the data
separate from the metadata. The \attr{href} attribute of
the {\elem{STREAM}} element, if present, provides the location of the data
in a URL-type syntax, for example:

\begin{plain}
\elemdef{STREAM}{ \attrval{href}{ftp://server.com/mydata.dat}\slash}

\par\elemdef{STREAM}{ \attrval{href}{ftp://server.com/mydata.dat}
        \attrval{expires}{2004-02-29T23:59:59}\slash}

\par\elemdef{STREAM}{ \attrval{href}{httpg://server.com/mydata.dat} 
        \attrval{actuate}{onLoad}\slash}

\par\elemdef{STREAM}{ \attrval{href}{file:///usr/home/me/mydata.dat}\slash}
\end{plain}

The examples are the well-known anonymous FTP and HTTP protocols.
\literalvalue{httpg} is an example of a Grid-based access to data through HTTPG;
finally, \literalvalue{file} is a reference to a local file.
VOTable parsers are not required to understand arbitrary protocols,
but are required to understand the  three common protocols
\literalvalue{file:}, \literalvalue{http:} and \literalvalue{ftp:}.

There are further attributes of the {\elem{STREAM}}
element that may be useful. The {\attr{expires}}
attribute indicates the expiration time of the data;
this is useful when data are dynamically created and stored 
on some staging disk where files only persist for a specified 
lifetime and are then automatically deleted.
The {\attr{expires}}
attribute expresses when a remote resource ceases to become valid,
and is expressed in Universal Time in the same way as the FITS
specification [2], itself conforming to the ISO 8601 standard.

The {\attr{rights}}
attribute expresses authentication information that may be necessary
to access the remote resource. If  the VOTable document is suitably
encrypted, this attribute could be used to store a password.

The {\attr{actuate}}
attribute is borrowed from the XML Xlink specification, expressing
when the remote link should be actuated. The default is {\literalvalue{onRequest}},
meaning that the data is only fetched when explicitly requested (like
a link on an HTML page), and the {\literalvalue{onLoad}}
value means that data should be fetched as soon as possible (like an
embedded image on an HTML page).

\section{Definitions of Primitive Datatypes}
\label{sec:datatypes}

This section describes the primitives summarized in 
\Tref{primitives}{the table of primitives}
and their representations in the \elem{BINARY}/\elem{BINARY2}  
and \elem{TABLEDATA} serializations 
(see \Aref{sec:data}{section 5}).
In the following, the term ``hexadigit'' designates the ASCII numbers
\literalvalue{0} to \literalvalue{9}, or the ASCII lower- or upper-case letters
\literalvalue{a} to \literalvalue{f} (i.e. a digit in a hexadecimal representation
of a number).

The representation of null values is discussed in \Aref{sec:NULL}{section 5.5}.

\begin{itemize}
\item {\bf Logical}\quad If the value of the {\attr{datatype}}
attribute specifies data type {\literalvalue{boolean}},
the contents of the field{ }shall consist of the \elem{BINARY}/\elem{BINARY2} serialization of
ASCII \literalvalue{T}, \literalvalue{t},  or \literalvalue{1} indicating true, and
ASCII \literalvalue{F}, \literalvalue{f}, or \literalvalue{0} indicating false.
The {\em null} value is indicated by an ascii NULL [0x00],
a space [0x20]
or a question mark \literalvalue{?} [0x3f].
The acceptable representations in the \elem{TABLEDATA} serialization
also include any capitalisation variation of the 
strings \literalvalue{true}  and \literalvalue{false} (e.g. \literalvalue{tRUe} or \literalvalue{FalsE}).

\item {\bf Bit Array} \quad If the value of the {\attr{datatype}}
attribute specifies data type {\literalvalue{bit}},
the contents of the field{ }in the \elem{BINARY}/\elem{BINARY2} serialization shall consist of
a sequence of bits starting with the most significant bit; the bits
following shall be in order of decreasing significance, ending with
the least significant bit. A bit field shall be composed of the
smallest number of bytes that can accommodate the number of elements
in the field. Padding bits shall be 0.
The representation of a bit array in the \elem{TABLEDATA} serialization
is made by a sequence of ASCII \literalvalue{0} and \literalvalue{1} characters.

\item {\bf Byte}\quad If the value of the {\attr{datatype}}
attribute specifies data type {\literalvalue{unsignedByte}},
the field shall contain in the \elem{BINARY}/\elem{BINARY2} serialization a byte 
(8-bits) representing a number in the
range 0 to 255. 
In the case of an array of bytes (\attrval{arraysize}{*}),
also known as a ``blob", the bytes are stored consecutively.
The representation of a byte in the \elem{TABLEDATA} serialization
can be its {\em decimal} representation (a number between {\tt0} and {\tt255})
or its {\em hexadecimal} representation when starting with {\tt0x} and 
followed by one or two hexadigits,
(e.g. {\tt0xff}), separated by at least one space from the next one
in the case of an array of bytes.

\item {\bf Character}\quad If the value of the {\attr{datatype}}
attribute specifies data type {\literalvalue{char}},
the field shall contain in the \elem{BINARY}/\elem{BINARY2} serialization an ASCII 
(7-bit) character. 
The \attr{arraysize} attribute
indicates a character string composed of ASCII text. 
The \elem{BINARY}/\elem{BINARY2} serialization follows the 
FITS rules for character strings,
and a character string may therefore be terminated by an ASCII 
NULL [0x00]
before the length specified in the \attr{arraysize} attribute.
In this case characters after the first ASCII NULL are not defined,
and a string having the number of characters identical to
the \attr{arraysize} value is not NULL terminated. 
Characters should be represented in the \elem{TABLEDATA} serialization
using the normal rules for encoding XML text: 
the ampersand (\&) can be written \verb+&amp;+ (symbolic representation)
or \verb+&#38;+ (decimal representation) or 
\verb+&#x26;+ (hexadecimal representation); the less-than ($<$) and greater-then ($>$) symbols should be coded \verb+&lt;+ and \verb+&gt;+ 
or \verb+&#x3C;+ and \verb+&#x3E;+.
Also note also the significance of the {\em white space} characters
in the \elem{TABLEDATA} serialization
(\Arefs{elem:TD}{section 5.1})

\item {\bf Unicode Character}\quad If the value of the {\attr{datatype}}
attribute specifies data type {\literalvalue{unicodeChar}},
the field shall contain a Unicode character.
The \attr{arraysize} attribute
indicates a string composed of Unicode text,
which enables representation of text in many non-Latin alphabets.
Each Unicode character is represented in the \elem{BINARY}/\elem{BINARY2} serialization by 
two bytes, using the big-endian UCS-2 encoding (ISO-10646-UCS-2).
The representation of a Unicode character in the  \elem{TABLEDATA} serialization
follows the XML specifications, 
and e.g. the Cyrillic uppercase ``Ya'' can be written 
\verb+&#x042F;+ in UTF-8.
Also note the significance of the {\em white space} characters
in the \elem{TABLEDATA} serialization
(\Arefs{elem:TD}{section 5.1})

\item {\bf 16-Bit Integer}\quad If the value of the {\attr{datatype}}
attribute specifies datatype {\literalvalue{short}}, 
the data in the \elem{BINARY}/\elem{BINARY2} serialization shall consist of
big-endian twos-complement signed 16-bit integers 
(the most significant byte first). 
The representation of a Short Integer in the \elem{TABLEDATA} serialization
is either its decimal representation between -32768 and 32767
   made of an optional {\tt-} or {\tt+} sign followed by digits,
   or its hexadecimal representation when starting with {\tt0x}
   and followed by 1 to 4 hexadigits.

\item {\bf 32-Bit Integer }\quad If the value of the {\attr{datatype}}
attribute specifies datatype {\literalvalue{int}},
the data in the \elem{BINARY}/\elem{BINARY2} serialization shall consist of 
big-endian twos-complement signed 32-bit
integer contained in four bytes, with the most significant first, 
and subsequent bytes in order of decreasing significance. 
  The representation of an Integer in the \elem{TABLEDATA} serialization
  is either its decimal representation between -2147483648 and 2147483647
  made of an optional {\tt-} or {\tt+} sign followed by digits,
  or its hexadecimal representation when starting with {\tt0x}
  and followed by 1 to 8 hexadigits;

\item {\bf 64-Bit Integer}\quad If the value of the {\attr{datatype}}
attribute specifies datatype {\literalvalue{long}},
the data in the \elem{BINARY}/\elem{BINARY2} serialization shall consist of 
big-endian twos-complement signed 64-bit integers
contained in eight bytes, with the most significant byte first,
and subsequent bytes in order of decreasing significance. 
The representation of a Long Integer in the \elem{TABLEDATA} serialization
  is either its decimal representation between -9223372036854775808
  and 9223372036854775807
  made of an optional {\tt-} or {\tt+} sign followed by digits,
  or its hexadecimal representation when starting with {\tt0x}
  and followed by 1 to 16 hexadigits;

\item {\bf Single Precision Floating Point}\quad If
the value of the {\attr{datatype}} attribute specifies datatype {\literalvalue{float}},
the data in the \elem{BINARY}/\elem{BINARY2} serialization shall consist of 
ANSI/IEEE-754 32-bit floating point numbers in big-endian order. 
All IEEE special values including NaN are recognized.
The representation of a Floating Point number in the 
\elem{TABLEDATA} serialization is made of an optional {\tt-} or {\tt+},
followed by the ASCII representation of a positive decimal number,
and followed eventually by the ASCII letter \literalvalue{E} or  \literalvalue{e}
introducing the base-10 exponent made of an optional {\tt-} or {\tt+}
followed by 1 or 2 digits. The number must be within the limits of the
IEEE floating-point definition (around $\pm3.4\cdot10^{38}$; numbers with
absolute value less than about $1.4\cdot10^{-45}$ are equated to zero).
The special
values \literalvalue{+Inf}, \literalvalue{-Inf}, and \literalvalue{NaN} are accepted.

\item {\bf Double Precision Floating Point}\quad If
the value of the {\attr{datatype}}
attribute specifies datatype {\literalvalue{double}},
the data in the \elem{BINARY}/\elem{BINARY2} serialization shall consist of ANSI/IEEE-754
64-bit double precision floating point numbers in big-endian order. 
All IEEE special values including NaN are recognized.
The representation of a Double number in the 
\elem{TABLEDATA} serialization is made of an optional {\tt-} or {\tt+},
followed by the ASCII representation of a positive decimal number,
and followed eventually by the ASCII letter \literalvalue{E} or  \literalvalue{e}
introducing the base-10 exponent made of an optional {\tt-} or {\tt+}
followed by 1 to 3 digits. The number must be within the limits of the
IEEE floating-point definition (around $\pm1.7\cdot10^{308}$; numbers with
absolute value less than about $5\cdot10^{-324}$ are equated to zero).
The special
values \literalvalue{+Inf}, \literalvalue{-Inf}, and \literalvalue{NaN} are accepted.

\item {\bf Single Precision Complex}\quad If the value of the {\attr{datatype}}
attribute specifies datatype {\literalvalue{floatComplex}},
the data in the \elem{BINARY}/\elem{BINARY2} serialization shall consist of a sequence of 
pairs of 32-bit single precision floating point numbers in big-endian order. 
The first member of each
pair shall represent the real part of a complex number and the
second member shall represent the imaginary part of that complex
number.
The representation of a Floating Complex number in the 
\elem{TABLEDATA} serialization is made of two representations
of a {\em  Single Precision Floating Point} numbers separated by 
whitespace, representing the real and imaginary part respectively.

\item {\bf Double Precision Complex}\quad If the
value of the {\attr{datatype}}
attribute specifies datatype {\literalvalue{doubleComplex}},
the data in the \elem{BINARY}/elem{BINARY2} serialization  shall consist of a
sequence of pairs of 64-bit double precision floating point numbers
in big-endian order.
The first member of each pair shall represent the real part of a
complex number and the second member of the pair shall represent the
imaginary part of that complex number.
The representation of a Double Complex number in the 
\elem{TABLEDATA} serialization is made of two representations
of a {\em  Double Precision Floating Point} numbers separated by 
whitespace, representing the real and imaginary part respectively.
\end{itemize}

\clearpage
\section{A Simplified View of the VOTable 1.3 Schema}
\label{dtd}
The XML Schema [6] defining a VOTable 1.3 document
is available from \ifhtx\else\\ \fi
\url{http://www.ivoa.net/xml/VOTable/v1.3}
In this section we illustrate this XML Schema
by a set of boxes describing the structure of a VOTable,
and the list of attributes of each VOTable element.

\subsection{Element Hierarchy}

The hierarchy of the elements existing in VOTable 1.3 is illustrated below;
it uses the following conventions:
\begin{itemize}
\item   {\em italicized} text represents {\em optional} elements;
\item   \order{} indicates that the order of the elements is mandatory, while
\item   \unorder{} {\em(open bullet)} indicates that the corresponding
        elements may occur in any order;
\item   \choice{} marks a choice between alternatives. 
\item   \deprecated{} marks a {\em deprecated} element (valid in Version 1.1,
        discouraged in Version 1.2 and later)
\item   $\cdots$ (dots) indicate that an element
        may be repeated. 
\item   \underline{underlined elements} may contain sub-elements,
        and are therefore explained in a dedicated box of the figure.
\end{itemize}

\ifhtx\begin{center}\Beg{tabular}{\bg{LavenderBlush} CELLPADDING=10}{||l||}
\else\bigskip
\begin{quote}\small
\fi

\def\element#1{\elemdef{#1}{}}
\def\optelem#1{{\fg{blue}$<$}{\footnotesize\em{\fg{DarkRed}#1}}{\fg{blue}$>$}}
\def\mv{\quad}
\def\paramonly{}

\noindent\begin{tabular}{cccc}
\begin{tabular}{|l|}\hline
\element{VOTABLE}			\\
\order \optelem{DESCRIPTION} 	\\
\deprecated {\small \optelem{COOSYS}$\cdots$} \\
\unorder \optelem{INFO}$\cdots$  \\
\unorder \optelem{PARAM}$\cdots$ \\
\unorder \optelem{GROUP}{\paramonly}$\cdots$ \\
\order \underline{\element{RESOURCE}}$\cdots$ 	\\
\order \optelem{INFO}$\cdots$  \\
\element{\slash VOTABLE}\\
\hline\end{tabular} 
&
\begin{tabular}{|l|}\hline
\element{RESOURCE}\\
\order \optelem{DESCRIPTION}	\\
\unorder \optelem{INFO}$\cdots$	\\
\deprecated \optelem{COOSYS}$\cdots$ 	\\
\unorder \underline{\optelem{GROUP}{\paramonly}}$\cdots$ \\
\unorder \underline{\optelem{PARAM}}$\cdots$	\\
\order {\optelem{LINK}}$\cdots$	\\
\order \underline{\optelem{TABLE}}$\cdots$	\\
\order \underline{\optelem{RESOURCE}}$\cdots$	\\
\order \optelem{INFO}$\cdots$  \\
\element{\slash RESOURCE}\\
\hline\end{tabular}
&
\begin{tabular}{|l|}\hline
\element{TABLE}	\\
\order \optelem{DESCRIPTION}	\\
\unorder    \underline{\optelem{FIELD}}$\cdots$	\\
\unorder    \underline{\optelem{PARAM}}$\cdots$	\\
\unorder    \underline{\optelem{GROUP}}$\cdots$	\\
\order \optelem{LINK}$\cdots$	\\
\order \underline{\optelem{DATA}}	\\
\order \optelem{INFO}$\cdots$ \\
\element{\slash TABLE}	\\
\hline\end{tabular}
&
\begin{tabular}{@{}c@{}}        
\begin{tabular}{|l|}\hline
\element{FIELD}	\\
\order \optelem{DESCRIPTION}	\\
\order \underline{\optelem{VALUES}}	\\
\order \optelem{LINK}$\cdots$	\\
\element{\slash FIELD}\\
\hline\end{tabular}
\\ \\
\begin{tabular}{|l|}\hline
\element{PARAM} \\
\order \optelem{DESCRIPTION}    \\
\order \underline{\optelem{VALUES}}     \\
\order \optelem{LINK}$\cdots$   \\
\element{\slash PARAM}\\
\hline\end{tabular}
\end{tabular}                   
\\
\begin{tabular}{|l|} \hline
  \element{DATA}     \\
    \choice {\element{TABLEDATA}} \\ 
        \mv\order \optelem{TR}$\cdots$ \\
        \mv\mv\order \element{TD}$\cdots$ \\
    \choice {\element{BINARY}} \\ 
        \mv\order \element{STREAM}\\
    \choice {\element{BINARY2}} \\
        \mv\order \element{STREAM}\\
    \choice {\element{FITS}}\\
        \mv\order \element{STREAM}\\
  \element{\slash DATA}\\
  \order \optelem{INFO}$\cdots$  \\
\hline\end{tabular}
&
\begin{tabular}{|l|}\hline
\element{GROUP} \\
\order \optelem{DESCRIPTION}	\\
\unorder    \optelem{FIELDref}$\cdots^{(t)}$ \\
\unorder    \underline{\optelem{PARAM}}$\cdots$	\\
\unorder    \optelem{PARAMref}$\cdots$	\\
\unorder    \underline{\optelem{GROUP}}$\cdots$	\\
\element{\slash GROUP}\\
\hline
{{\footnotesize\em(t) only within \element{TABLE}}}\\
\hline\end{tabular}
&
\begin{tabular}{|l|}\hline
\element{VALUES}	\\
\order \optelem{MIN}	\\
\order \optelem{MAX}	\\
\order \optelem{OPTION}$\cdots$	\\
\mv\unorder\optelem{OPTION}$\cdots$\\
\element{\slash VALUES}\\
\hline\end{tabular}
\\ \\
\\
\end{tabular}

\ifhtx\End{tabular}\end{center}\else
\end{quote}
\fi

\subsection{Attribute Summary}
The list of the attributes is summarized in the table below, 
with the following conventions:
\begin{itemize}
\item   Attributes written in bold are \requiredattr{required attributes}
\item   Attributes written in a {fixed font} are \attr{optional}.
\item   Attributes written in {\it italics}
        are not part of VOTable 1.3, but are {\it reserved}
        for possible extensions (mentioned in an Appendix).
\end{itemize}

\ifhtx\begin{center}\Beg{tabular}{\bg{LavenderBlush} CELLPADDING=4}{||l||}
\else\bigskip
\begin{quote}\small
\fi
\def\attrx#1{{\em\fg{DarkRed}#1}}
\begin{tabular}{cccccc}
\begin{tabular}{|l|}\hline
\multicolumn{1}{|c|}{\elem{VOTABLE}}\\
\multicolumn{1}{|c|}{{\em(\Aref{elem:VOTABLE}{definition})}}\\ \hline{}
   \attr{ID}\\		
   \attr{version}\\
\hline\end{tabular}
&
\begin{tabular}{|l|}\hline
\multicolumn{1}{|c|}{\elem{RESOURCE}}\\
\multicolumn{1}{|c|}{{\em(\Aref{elem:RESOURCE}{definition})}}\\ \hline{}
   \attr{ID}\\
   \attr{name}\\
   \attr{type}\\
   \attr{utype}\\
\hline\end{tabular}
&
\begin{tabular}{|l|}\hline
\multicolumn{1}{|c|}{\elem{TABLE}}\\
\multicolumn{1}{|c|}{{\em(\Aref{elem:TABLE}{definition})}}\\ \hline{}
   \attr{ID}\\
   \attr{name}\\
   \attr{ucd}\\
   \attr{utype}\\
   \attr{ref}\\
   \attr{nrows}\\
\hline\end{tabular}
&
\begin{tabular}{|l|}\hline
\multicolumn{1}{|c|}{\elem{INFO}}\\
\multicolumn{1}{|c|}{{\em(\Aref{elem:INFO}{definition})}}\\ \hline{}
   \attr{ID}\\
   \requiredattr{name}\\
   \requiredattr{value}\\
   \attr{xtype}\\
   \attr{ref}\\
   \attr{unit}\\
   \attr{ucd}\\
   \attr{utype}\\
\hline\end{tabular}
&
\begin{tabular}{|l|}\hline
\multicolumn{1}{|c|}{\elem{STREAM}}\\
\multicolumn{1}{|c|}{{\em(\Aref{elem:STREAM}{definition})}}\\ \hline{}
   \attr{type}\\
   \attr{href}\\
   \attr{actuate}\\
   \attr{encoding}\\
   \attr{expires}\\
   \attr{rights}\\
\hline\end{tabular}
&
\begin{tabular}{@{}c@{}}	
\begin{tabular}{|l|}\hline
\multicolumn{1}{|c|}{\elem{FITS}}\\
\multicolumn{1}{|c|}{{\em(\Aref{elem:FITS}{definition})}}\\ \hline{}
   \attr{extnum}\\
\hline\end{tabular}
	\\ \\
\begin{tabular}{|l|}\hline
\multicolumn{1}{|c|}{\elem{TR}}\\
\multicolumn{1}{|c|}{{\em(\Aref{elem:TR}{definition})}}\\ \hline{}
     \attrx{ID}\\
\hline\end{tabular}
	\\ \\
\begin{tabular}{|l|}\hline
\multicolumn{1}{|c|}{\elem{TD}}\\
\multicolumn{1}{|c|}{{\em(\Aref{elem:TD}{definition})}}\\ \hline{}
   \attrx{encoding}\\
\hline\end{tabular}
\end{tabular}			
\\ \\
\\ \\
%
%
\begin{tabular}{@{}c@{}}	
\begin{tabular}{|l|}\hline
\multicolumn{1}{|c|}{\elem{GROUP}}\\
\multicolumn{1}{|c|}{{\em(\Aref{elem:GROUP}{definition})}}\\ \hline{}
   \attr{ID}\\
   \attr{name}\\
   \attr{ref}\\
   \attr{ucd}\\
   \attr{utype}\\
\hline\end{tabular}
\\ \\
\end{tabular}			
&
\begin{tabular}{|l|}\hline
\multicolumn{1}{|c|}{\elem{PARAM}}\\
\multicolumn{1}{|c|}{{\em(\Aref{elem:PARAM}{definition})}}\\ \hline{}
   \attr{ID}\\
   \attr{unit}\\
   \requiredattr{datatype}\\
   \attr{precision}\\
   \attr{width}\\
   \attr{xtype}\\
   \attr{ref}\\
   \requiredattr{name}\\
   \attr{ucd}\\
   \attr{utype}\\
   \attr{arraysize}\\
   \requiredattr{value}\\
\hline\end{tabular}
&
\begin{tabular}{|l|}\hline
\multicolumn{1}{|c|}{\elem{FIELD}}\\
\multicolumn{1}{|c|}{{\em(\Aref{elem:FIELD}{definition})}}\\ \hline{}
   \attr{ID}\\
   \attr{unit}\\
   \requiredattr{datatype}\\
   \attr{precision}\\
   \attr{width}\\
   \attr{xtype}\\
   \attr{ref}\\
   \requiredattr{name}\\
   \attr{ucd}\\
   \attr{utype}\\
   \attr{arraysize}\\
   \attrx{type}\\
\hline\end{tabular}
&
\begin{tabular}{@{}c@{}}	
\begin{tabular}{|l|}\hline
\multicolumn{1}{|c|}{\elem{FIELDref}}\\
\multicolumn{1}{|c|}{{\em(\Aref{elem:FIELDref}{definition})}}\\ 
   \hline{}
   \requiredattr{ref}\\
   \attr{ucd}\\
   \attr{utype}\\
\hline\end{tabular}
\\ \\
\\ \\
\begin{tabular}{|l|}\hline
\multicolumn{1}{|c|}{\elem{PARAMref}}\\
\multicolumn{1}{|c|}{{\em(\Aref{elem:PARAMref}{definition})}}\\ 
   \hline{}
   \requiredattr{ref}\\
   \attr{ucd}\\
   \attr{utype}\\
\hline\end{tabular}
\end{tabular}			
&
\begin{tabular}{@{}c@{}}
\begin{tabular}{|l|}\hline
\multicolumn{1}{|c|}{\elem{MIN}}\\
\multicolumn{1}{|c|}{{\em(\Aref{elem:MIN}{definition})}}\\ \hline{}
   \requiredattr{value}\\
   \attr{inclusive}\\
\hline\end{tabular}
	\\ \\
\begin{tabular}{|l|}\hline
\multicolumn{1}{|c|}{\elem{MAX}}\\
\multicolumn{1}{|c|}{{\em(\Aref{elem:MAX}{definition})}}\\ \hline{}
   \requiredattr{value}\\
   \attr{inclusive}\\
\hline\end{tabular}
	\\ \\
\begin{tabular}{|l|}\hline
\multicolumn{1}{|c|}{\elem{OPTION}}\\
\multicolumn{1}{|c|}{{\em(\Aref{elem:OPTION}{definition})}}\\ \hline{}
   \attr{name}\\
   \requiredattr{value}\\
\hline\end{tabular}
\end{tabular}
&
\begin{tabular}{@{}c@{}}	
\begin{tabular}{|l|}\hline
\multicolumn{1}{|c|}{\elem{VALUES}}\\
\multicolumn{1}{|c|}{{\em(\Aref{elem:VALUES}{definition})}}\\ \hline{}
   \attr{ID}\\
   \attr{type}\\
   \attr{null}\\
   \attr{ref}\\
\hline\end{tabular}
\\ \\
\begin{tabular}{|l|}\hline
\multicolumn{1}{|c|}{\elem{LINK}}\\
\multicolumn{1}{|c|}{{\em(\Aref{elem:LINK}{definition})}}\\ \hline{}
   \attr{ID}\\
   \attr{content-role}\\
   \attr{content-type}\\
   \attr{title}\\
   \attr{value}\\
   \attr{href}\\
   \attrx{action}\\
\hline\end{tabular}
\end{tabular}			
\\
%
%

%

\end{tabular}

\ifhtx\End{tabular}\end{center}\else\end{quote}
\fi

\section{MIME Type}
\label{sec:mime}
A VOTable document should be introduced by a 
Multipurpose Internet Mail Extensions media type, or MIME type.
MIME type syntax is described in RFC 2045 section 5.1, and
its semantics in RFC 2046.
Associating a MIME type to a document enables the {\em data consumer}
(an application or a web browser) to launch the desired application
({\em e.g.} a visualisation tool).

In the HTTP protocol (RFC 2616), the MIME type is the value specified by the
{\sf Content-Type:} header. The recommended MIME type describing
a VOTable document is {\literalvalue{\sf application/x-votable+xml}}:
the {\bf x-} prefix indicates an experimental type, and
is required for non-registered media types; and the 
{\bf+xml} suffix (defined by RFC 3023 section 7)
indicates that the type describes a specialization of XML.
This type may be accompanied by an optional parameter
{\literalvalue{\sf serialization}}
with a value specifying the serialization type used for table data within the
document, one of TABLEDATA, FITS, BINARY or BINARY2,
interpreted case-insensitively.
In the absence of this parameter, any of the serializations may be
encountered.  If multiple different serializations are used in the same
document, this parameter must not be used.

Alternatively the {\literalvalue{\sf text/xml}} MIME type is acceptable
for services delivering data which are expected to be
visualized by humans in a browser; this MIME type 
would preferably be associated with an XSL style sheet,
for a presentation of well-formatted tables. It is expected
that a few typical XSL style sheets will be accessible from
the IVOA site.
Note that use of the {\em text} top-level media type means that
line breaks must be represented as a CRLF sequence
(RFC 2046, section 4.1.1).

For both of these MIME types, RFC 3023 also defines the optional
parameter {\literalvalue{\sf charset}}.
If this parameter is not supplied, US-ASCII is assumed.

Any of the following Content-Type header values may therefore be used
by a service producing VOTables with the TABLEDATA serialization:
\begin{itemize}
  \item {\sf text/xml}
  \item {\sf text/xml; charset="iso-8859-1"}
  \item {\sf application/x-votable+xml}
  \item {\sf application/x-votable+xml; serialization=tabledata}
  \item {\sf application/x-votable+xml; serialization=TABLEDATA; charset=iso-8859-1}
\end{itemize}

\section{Version History}
\label{diff}

\subsection{Differences Between Versions 1.1 and 1.2}
\label{diff1.1-1.2}
The differences between version 1.2 of VOTable and the preceding
version 1.1 are:

\ifhtx\Beg{tabular}{\bg{LavenderBlush} CELLPADDING=4}{||p||}
\fi
\begin{itemize}
\item   \elem{COOSYS} is deprecated, in favor of a reference
        to the {\em Space-Time Coordinate} (STC) data model
        (see \Arefs{sec:utype}{utype} and the IVOA note
        {\it Referencing STC in VOTable}[7])
\item  \elem{GROUP} may appear as a direct child of
	\elem{VOTABLE} and \elem{RESOURCE} (where {\em COOSYS} was 
	acceptable)
\item  The usage of UCD1+ is recommended (\Arefs{sec:ucd}{section 4.5})
\item  The \attr{xtype} attribute was added 
	(see  \Arefs{sec:xtype}{section 4.3})
\item  The {\elem{INFO}} element (\Arefs{elem:INFO}{INFO}) is made 
	more similar
        to the \elem{PARAM} element, but with \attrval{datatype}{char}
        and \attrval{arraysize}{*} (i.e. is a {\em String});
	it may have attributes \attr{utype},  \attr{ucd},
		\attr{ref}, \attr{unit}
\item  The {\elem{INFO}} element may occur before the closing
        tags \elem{/TABLE} and \elem{/RESOURCE} 
        and \elem{/VOTABLE}
        (enables {\em post-operational diagnostics})
\item  The {\elem{FIELDref}} and {\elem{PARAMref}} elements may have
	a \attr{utype} and \attr{ucd} attribute.
\item  Naming conventions of \elem{GROUP} elements which specify some
        properties of a relational schema 
	(see \Arefs{sec:relation}{section 4.10}).
\item   The recommended and acceptable mime types have been made explicit
	(\Arefs{sec:mime}{section 7.3})
\item   The representation of arrays in cells has been made explicit 
	(\Arefs{sec:dim}{section 2.2})
\item   Detailed and clarified the conventions and recommendations concerning
	\attr{name}, \attr{ID} and \attr{ref} attributes
\item	Appendix A7 was a proposition for additional \attr{utype} 
        attributes in groups and tables; it is now included in VOTable 1.2.
	\Arefx{sec:addesc}{Appendix A7} now contains a new proposal 
	(May/June 2009) for multiple descriptions and titles.
\end{itemize}
\ifhtx\End{tabular}\fi

\subsection{Differences Between Versions 1.2 and 1.3}
\label{diff1.2-1.3}
The differences between version 1.3 of VOTable and the preceding
version 1.2 are:

\ifhtx\Beg{tabular}{\bg{LavenderBlush} CELLPADDING=4}{||p||}
\fi
\begin{itemize}
\item The \elem{BINARY2} serialization has been introduced
      (\Aref{sec:BIN2}{section 5.4}).
      \elem{BINARY} is mildly deprecated.
\item The usage and semantics of an empty \elem{TD} element in the
      \elem{TABLEDATA} serialization have changed.
      In VOTable 1.3, an empty \elem{TD} element is legal for any datatype
      (previously it was illegal for integer types)
      and it always denotes a null value
      (previously it indicated NaN for floating point types).
      This change means that the different serializations no longer have
      exactly the same capabilities for data representation.
\item A new \Aref{sec:NULL}{section 5.5} has been added to clarify
      usage and encoding for null values.
\item In view of the new options for flagging null values introduced by the
      BINARY2 and TABLEDATA changes above, use of the
      \elem{VALUES} \attr{null} attribute is now deprecated in most cases.
      It is in any case explicitly deprecated for floating point values,
      in favour of NaN.
\item The description of the \elem{LINK} element (\Aref{sec:link}{section 3.5})
      has been clarified, and a new \attrval{content-role}{type}
      example added, with discussion of its application to SKOS concepts.
\item The schema datatype declaration of the \elem{LINK} \attr{content-type}
      attribute has been changed from {\tt NMTOKEN} to {\tt token}.
      The {\tt NMTOKEN} datatype in VOTable 1.2 was a mistake, since it
      would not have permitted the MIME type example in the text.
\item The MIME type section (now \Aref{sec:mime}{section 8}) now describes
      the new {\tt serialization} parameter that can be used to specify
      serialization type.
\item The representation of STC information in \Aref{example1}{section 3.1}
      and \Aref{query}{appendix A.2}
      has been modified to reflect the recommended usage from the
      {\em STC in VOTable} Note [7].  This usage is recommended even for
      VOTable 1.2, so this change to the VOTable document represents
      an update of advice rather than a change to the normative part
      of the VOTable standard.  Additionally, text has been added
      encouraging declaration of the STC metadata where possible.
\item A new \Aref{sec:voarch}{section 1.4} has been added explaining
      the place of VOTable in the IVOA Architecture.
\end{itemize}
\ifhtx\End{tabular}\fi

\section{References}
\noindent [1] Accomazzi {\it  et. al, Describing Astronomical Catalogues and
Query Results with XML
}\\ \hspace*{2em}{\url{http://cds.u-strasbg.fr/doc/astrores.htx}}

\noindent [2] {\it FITS: Flexible Image Transport
Specification}, specifically the Binary Tables
Extension
\\ \hspace*{2em}{\url{http://fits.gsfc.nasa.gov/}}

\noindent [3] {\it Standards for Astronomical
Catalogues: Units, CDS Strasbourg}
\\ \hspace*{2em}{\url{http://cdsarc.u-strasbg.fr/doc/catstd-3.2.htx}}
\\ \hspace*{2em} {\it See also Section 4 in} {Greisen and Calabretta} 2002,
        A\&A {\bf 395}, 1061; and the
	 \ifhtx\else\\ \hspace*{2em}\fi
	IAU Recommendations concerning Units
        from the {\em IAU Style Manual} by  G.A. Wilkins (1989)
        available at \ifhtx\else\\ \hspace*{2em}\fi
        \url{http://www.iau.org/science/publications/proceedings_rules/units/}

\noindent [4] {\it Unified Content
Descriptors}
\\ \hspace*{2em}{\url{http://cds.u-strasbg.fr/doc/UCD.htx}} (UCD1)
\\ \hspace*{2em}{\url{http://www.ivoa.net/twiki/bin/view/IVOA/IvoaUCD}}

\noindent [5] {\it ASU: Astronomical Server URL, CDS
Strasbourg}
\\ \hspace*{2em}{\url{http://cds.u-strasbg.fr/doc/asu.html}}

\noindent [6] {\it XML Schema: W3C Document}
\\ \hspace*{2em}{\url{http://www.w3.org/XML/Schema}}

\noindent [7] {\it Referencing STC in VOTable}
\\ \hspace*{2em}{\url{http://ivoa.net/Documents/latest/VOTableSTC.html}}

\noindent [8] Arnold Rots
        {\it Space-Time Coordinate Metadata for the Virtual Observatory (v1.30)}
\\ \hspace*{2em}{\url{http://ivoa.net/Documents/latest/STC.html}}

\noindent [9] Arnold Rots
        {\it STC-S: Space-Time Coordinate (STC) Metadata Linear String Implementation}
\\ \hspace*{2em}{\url{http://www.ivoa.net/Documents/latest/STC-S.html}}

\noindent [10] {\it Registry of FITS conventions}
\\ \hspace*{2em}{\url{http://fits.gsfc.nasa.gov/fits\_registry.html}}

\noindent [11] {\it Table Access Protocol}
\\ \hspace*{2em}{\url{http://ivoa.net/Documents/latest/TAP.html}}
\par
\noindent [12] {\it IVOA Astronomical Data Query Language}
\\ \hspace*{2em}{\url{http://ivoa.net/Documents/latest/ADQL.html}}


\ifhtx\par\thickrule\par\Beg{tabular}{\bg{LightBlue} CELLPADDING=5}{l}\quad
\else\clearpage\fi
\appendix
\noindent {\bf\LARGE Appendices}
\ifhtx\End{tabular}\fi

\bigskip

\section{Possible VOTable extensions}
The definitions enclosed in this appendix
are {\bf not} part of VOTable 1.1, but are considered as candidates
for VOTable improvements. 

\subsection{VOTable LINK substitutions}
\label{LINK}

\begin{quote}\em \fg{DarkBlue}
  The \elem{LINK} element in Astrores [1]
  contains a mechanism for string substitution,
  which is a powerful way of defining a link to external data
  which adapts to each record contained in the table \elem{DATA}.
\end{quote}

When a {\elem{LINK}} element appears within a \elem{RESOURCE} or a
{\elem{TABLE}} element,
extra functionality is implied: the {\attr{href}}
attribute may not be a simple link, but instead 
a template for a link. If, in the  example of 
\Aref{example1}{myFavouriteGalaxies}, we add the link

\begin{verbatim}
  <LINK href="http://ivoa.net/lookup?Galaxy=${Name}&amp;RA=${RA}&amp;DE=${DE}"/>
\end{verbatim}

\noindent a substitution filter is applied in the context of a particular row.
For the first row of the table, the substitution would result in the URL

\begin{verbatim}
   http://ivoa.net/lookup?Galaxy=N%20224&RA=010.68&DE=%2b41.27
\end{verbatim}

Whenever the pattern {\tt{\$\{...\}}}
is found in the original link, the part in the braces is compared
with the set of {\attr{ID}} (preferably) or \attr{name}
attributes of the fields of the table. If a match is found, then the
value from that field of the selected row is used in place of the
{\tt{\$\{...\}}}. If no match is found, no substitution is made. Thus the
parser makes available to the calling application a value of the {\attr{href}}
attribute that depends on which row of the table has been selected.
Another way to think of it is that there is not a single link
associated with the table, but rather an implicitly defined new
column of the table. This mechanism can be used to connect each row
of the table to further information resources.


The purpose of the link is defined by the {\attr{content-role}}
attribute. The allowed values are {\literalvalue{query}} 
(see \Aref{query}{query mechanism}), 
{\literalvalue{hints}} for information for use by the application,
and {\literalvalue{doc}} for  human-readable documentation.

The column names invoked in the pattern of the \attr{href} attribute
of the \elem{LINK} element should exist in the document to 
generate meaningful links. 
In the common case where the VOTable was generated from a query
of a database and contains only some of the columns in that
database, it might be necessary to include columns additional to
those requested in order to ensure that the LINKS in the VOTable
are operational.
Such a \elem{FIELD} included ``by necessity'' is marked with 
the attribute \attrval{type}{hidden}. The primary key of
a relational table is a typical example of a \elem{FIELD} 
which would carry the \attrval{type}{hidden} attribute.

\subsection{VOTable Query Extension}
\label{query}

\begin{quote}\em\fg{DarkBlue}
  The metadata part included in  a \elem{RESOURCE} contains
  all the details necessary to create a {\em form} for querying
  the resource. The addition of a link having the \attr{action} 
  attribute can turn VOTable into a powerful query interface.
\end{quote}

\noindent In Astrores [1], the details on the input parameters available in
queries are described by the 
{\elem{PARAM}} and {\elem{FIELD}} elements, and the syntax used
to generate the actual query is described in the ASU [5] procotol:
the {\elem{FIELD}} or \elem{PARAM} elements are
paired in the form {\it name}{{\tt=}}{\it value},
where {\it name} is the contents of the
\attr{name} attribute of a \elem{FIELD} or \elem{PARAM}, 
and  {\it value} represents a constraint
written with the ASU conventions (e.g. \literalvalue{$<8$}
 or {\literalvalue{12.0..12.5}}
which denotes a range of values). 
Such pairs are  appended to the
{\attr{action}} specified in the {\elem{LINK}}
element contained in the {{\elem{RESOURCE}}},
separated by the ampersand (\&) symbol --
in a way quite similar to the HTML syntax used to 
describe a {\elem{FORM}}.

A special \attrval{type}{no\_query} attribute of the
\elem{PARAM} or \elem{FIELD} elements marks the fields
which are {\em not} part of the form, i.e. are ignored 
in the collection of {\it name}{{\tt=}}{\it value} pairs.

The following is an example of a transformation of the VOTable
in \Aref{example1}{the example} into a form interface:
\label{form1}
\ifhtx\Beg{tabular}{\bg{LightCyan} CELLPADDING=5}{||l||}
\else\begingroup\small
\fi
\begin{verbatim}
<?xml version="1.0"?>
<VOTABLE version="1.3" xmlns:xsi="http://www.w3.org/2001/XMLSchema-instance"
 xmlns="http://www.ivoa.net/xml/VOTable/v1.3"
 xmlns:stc="http://www.ivoa.net/xml/STC/v1.30" >
  <RESOURCE name="myFavouriteGalaxies" type="meta">
    <TABLE name="results">
      <DESCRIPTION>Velocities and Distance estimations</DESCRIPTION>
      <GROUP utype="stc:CatalogEntryLocation">
        <PARAM name="href" datatype="char" arraysize="*"
               utype="stc:AstroCoordSystem.href" value="ivo://STClib/CoordSys#UTC-ICRS-TOPO"/>
        <PARAM name="URI" datatype="char" arraysize="*"
               utype="stc:DataModel.URI" value="http://www.ivoa.net/xml/STC/stc-v1.30.xsd"/>
        <FIELDref utype="stc:AstroCoords.Position2D.Value2.C1" ref="col1"/>
        <FIELDref utype="stc:AstroCoords.Position2D.Value2.C2" ref="col2"/>
      </GROUP>
      <PARAM name="-out.max" ucd="meta.number" datatype="int" value="50">
        <DESCRIPTION>Maximal number of records to retrieve</DESCRIPTION>
      </PARAM>
      <FIELD name="RA"   ID="col1" ucd="pos.eq.ra;meta.main"
             datatype="float" width="6" precision="2" unit="deg"/>
      <FIELD name="Dec"  ID="col2" ucd="pos.eq.dec;meta.main"
             datatype="float" width="6" precision="2" unit="deg"/>
      <FIELD name="Name" ID="col3" ucd="meta.id;meta.main"
             datatype="char" arraysize="8*"/>
      <FIELD name="RVel" ID="col4" ucd="spect.dopplerVeloc" datatype="int"
             width="5" unit="km/s"/>
      <FIELD name="e_RVel" ID="col5" ucd="stat.error;spect.dopplerVeloc"
             datatype="int" width="3" unit="km/s"/>
      <FIELD name="R" ID="col6" ucd="pos.distance;pos.heliocentric"
             datatype="float" width="4" precision="1" unit="Mpc">
        <DESCRIPTION>Distance of Galaxy, assuming H=75km/s/Mpc</DESCRIPTION>
      </FIELD>
      <LINK content-role="query" action="myQuery?-source=myGalaxies&amp;" />
    </TABLE>
  </RESOURCE>
</VOTABLE>
\end{verbatim}
\ifhtx\End{tabular}
\else
\endgroup
\fi

\noindent Note that the {\elem{RESOURCE}} displaying the parameters accessible 
for a query has the {\attrval{type}{meta}}
attribute; it is also assumed that only one {\elem{LINK}}
having the {\attrval{content-role}{query}}
attribute together with an {\attr{action}}
attribute exists within the current {\elem{RESOURCE}}.
The \elem{PARAM} with \attrval{name}{-out.max} has been added in this
example to control the size of the result.

A valid query generated by this VOTable could be:

\begin{verbatim}
  myQuery?-source=myGalaxies&-out.max=50&R=10..100
\end{verbatim}


\subsection{Arrays of Variable-Length Strings}
\label{sec:arraystring}
Following the FITS conventions, strings are defined as arrays of
characters. This definition raises problems for the definition
of arrays of strings, which have then to be defined as 2D-arrays
of characters -- but in this case only the slowest-varying dimension
(i.e. the number of strings) can be variable. 
This limitation becomes severe when a table column contains a set
of remarks, each being made of a variable number of characters as 
occurs in practice.

FITS invented the {\em Substring Array} convention (defined in an appendix,
i.e. not officially approved) which defines a {\em separator} character
used to denote the end of a string and the beginning of the next one.
In this convention ($r${\tt A:SSTR}$w$/$ccc$) the total size of the character
array is specified by $r$, $w$ defines the maximum length of one string,
and $ccc$ defines the separator character as its ASCII equivalent value.
The possible values for the separator includes the space and any printable
character, but excludes the control characters.

Such arrays of variable-length strings are frequently useful e.g.
to enumerate a list of properties of an observed source, each property being
represented by a variable-length string.
A convention similar to the FITS one could be introduced in 
VOTable in the \attr{arraysize}
attribute, using the {\bf s} followed by the separator character;
an example can be \attrval{arraysize}{100s,}
indicating a string made of up to 100 characters, where the comma
is used to separate the elements of the array.

\subsection{FIELDs as Data Pointers}
\label{location}

Rather than requiring that all data described in the set of \elem{FIELD}s
are contained in a single stream which follows the metadata part, 
it would be possible to let the \elem{FIELD} act as 
a {\em pointer} to the actual data, either in the form of a URI or of
a reference to a component of a multipart document.

Each component of the data described by a \elem{FIELD} may effectively
have different requirements: while text data or small lists of numbers
are quite efficiently represented in pure XML, long lists like spectra
or images generate poor performances if these are converted to XML.
The method available to gain efficiency is to use a
binary representation of the {\em whole data stream} by means of the
\elem{STREAM} element -- at the price of delivering data in a totally non-human
readable format.

The following options would allow more flexibility in the way the 
various \elem{FIELD}s can be accessed:

\begin{itemize}
\item   a \elem{FIELD} can be declared as being a {\em pointer}
        with the addition of a \attrval{type}{location} value,
        meaning that the field contains a way to access the data, 
        and not the actual data;
\item   a \elem{FIELD} can contain a \elem{LINK} element marked 
        \attrval{type}{location} which contains in its
        \attr{href} attribute the partial URI to which the contents
        of the column cell is appended in order to generate a
        fully qualified URI.
\end{itemize}
Note that the \elem{LINK} is not required -- a \elem{FIELD} declared
with \attrval{type}{location} and containing no \elem{LINK} element
is assumed to contain URIs.

An example of a table describing a set of spectra could look like the following:

\ifhtx\Beg{tabular}{\bg{LightCyan} CELLPADDING=5}{||l||}
\else\small\fi
\begin{verbatim}
<TABLE name="SpectroLog">
  <FIELD name="Target" ucd="meta.id" datatype="char" arraysize="30*"/>
  <FIELD name="Instr" ucd="instr.setup" datatype="char" arraysize="5*"/>
  <FIELD name="Dur" ucd="time.expo" datatype="int" width="5" unit="s"/>
  <FIELD name="Spectrum" ucd="meta.ref.url" datatype="float" arraysize="*"
         unit="mW/m2/nm" type="location">
    <DESCRIPTION>Spectrum absolutely calibrated</DESCRIPTION>
    <LINK type="location" 
        href="http://ivoa.spectr/server?obsno="/>
  </FIELD>
  <DATA><TABLEDATA>
    <TR><TD>NGC6543</TD><TD>SWS06</TD><TD>2028</TD><TD>01301903</TD></TR>
    <TR><TD>NGC6543</TD><TD>SWS07</TD><TD>2544</TD><TD>01302004</TD></TR>
  </TABLEDATA></DATA>
</TABLE>
\end{verbatim}\ifhtx\End{tabular}\else\normalsize\fi

\noindent
The reading program has therefore to retrieve the data 
for this first row by resolving the URI
\begin{plain}
{\tt http://ivoa.spectr/server?obsno=01301903}
\end{plain}

\noindent
The same method could also be immediately applicable to  {\em Content-ID}s
which designate elements of a multipart message, using the protocol
prefix {\tt cid:} [RFC2111]

Note that the {\em VOTable LINK substitution} proposed in 
\Aref{LINK}{Appendix A} fills a similar functionality: 
generate a pointer which can incorporate in its address components
from the \elem{DATA} part for the VOTable.

\subsection{Encoding Individual Table Cells}
\label{sec:b64}
Accessing binary data improves quite significantly the efficiency
both in storage and CPU usage, especially when one compares with the
XML-encoded data stream. But binary data cannot be included in the
same stream as the metadata description, unless a dedicated coding
filter is applied which converts the binary data into an ASCII representation.
The base64 is the most commonly used filter for this conversion, where 
3 bytes of data are coded as 4 ASCII characters, which implies an overhead of
33\% in storage, and some (small) computing time necessary for the reverse 
transformation.

In order to keep the full VOTable document in a unique stream,
VOTable 1.0 introduced the \attr{encoding} attribute in the
\elem{STREAM} element, meaning that the data, stored as binary records,
are converted into some ASCII representation compatible with the 
XML definitions. One drawback of this method is that the entire data
contents become non human-readable.

The addition of the \attr{encoding} attribute in the \elem{TD} element
allows the data server to decide, at the cell level, whether it is more
efficient to distribute the data as binary-encoded or as edited
values. The result may look like the following:

\ifhtx\Beg{tabular}{\bg{LightCyan} CELLPADDING=5}{||l||}\fi
\begin{verbatim}
<TABLE name="SpectroLog">
  <FIELD name="Target" ucd="meta.id" datatype="char" arraysize="30*"/>
  <FIELD name="Instr" ucd="instr.setup" datatype="char" arraysize="5*"/>
  <FIELD name="Dur" ucd="time.expo" datatype="int" width="5" unit="s"/>
  <FIELD name="Spectrum" ucd="phot.flux;em.opt" datatype="float" arraysize="*"
         unit="mW/m2/nm" precision="E3"/>
  <DATA><TABLEDATA>
    <TR><TD>NGC6543</TD><TD>SWS06</TD><TD>2028</TD><TD encoding="base64">
    QJKPXECHvndAgMScQHul40CSLQ5ArocrQLxiTkC3XClAq0OWQKQIMUCblYFAh753QGij10BT
    Em9ARKwIQExqf0BqbphAieuFQJS0OUCJWBBAhcrBQJMzM0CmRaJAuRaHQLWZmkCyhytAunbJ
    QLN87kC26XlA1KwIQOu+d0DsWh1A5an8QN0m6UDOVgRAxO2RQM9Lx0Din75A3o9cQMPfO0C/
    dLxAvUeuQKN87kCXQ5ZAjFodQH0vG0B/jVBAgaHLQI7Ag0CiyLRAqBBiQLaXjUDYcrBA8p++
    QPcKPUDg7ZFAwcKPQLafvkDDlYFA1T99QM2BBkCs3S9AjLxqQISDEkCO6XlAmlYEQKibpkC5
    wo9AvKPXQLGBBkCs9cNAuGp/QL0euEC4crBAuR64QL6PXEDOTdNA2987QN9T+EDoMSdA8mZm
    QOZumEDDZFpAmmZmQGlYEEBa4UhAivGqQLel40Dgan9A4WBCQLNcKUCIKPZAk1P4QNWRaEEP
    kWhBKaHLQTkOVkFEan9BUWBCQVyfvg==
    </TD></TR>
  </TABLEDATA></DATA>
</TABLE>
\end{verbatim}
\ifhtx\End{tabular}\fi
\par

\noindent 
When decoded, the contents of the last column is the binary representation
of the spectrum, as defined in \Aref{sec:BIN}{the BINARY serialization};
no length prefix is required here, the total length of the array being
implicitly defined by the length of the encoded text.

\subsection{Very Large Arrays}
The \elem{BINARY} and \elem{BINARY2} serializations of variable-length arrays 
(\Aref{sec:BIN}{section 5.3}, \Aref{sec:BIN2}{5.4}) uses a 4-byte prefix containg the number of
items of the array. This convention imposes an absolute maximal
number of $2^{31}-1$ elements. This limit could be releaved 
with a new \attr{arrayprefix} attribute.

\subsection{Additional Descriptions and Titles}
\label{sec:addesc}
The same table may be used in several contexts, and it was for
instance expressed a wish to include in \elem{TABLE}  and
\elem{FIELD} descriptions and titles (captions) in a form
suitable for a publication (latex)
in addition to the ascii-only descriptions currently acceptable.
The following example is an illustration of this extension:
\begin{verbatim}
<TABLE name="Model_A">
  <DESCRIPTION>Star luminosities in Model A</DESCRIPTION>
  <DESCRIPTION context="latex">$L(T_{eff})$ in Model {\bf A}</DESCRIPTION>
  <FIELD name="Teff" datatype="float" unit="K" ucd="phys.temperature.effective">
     <DESCRIPTION>Effective temperature</DESCRIPTION>
     <TITLE context="latex">$T_{eff}$</TITLE>
  </FIELD>
  <FIELD name="Lum" datatype="float" unit="Lsun" ucd="phys.luminosity">
     <DESCRIPTION>Corresponding luminosity in Model A</DESCRIPTION>
     <DESCRIPTION context="latex">$L(T_{eff})$</DESCRIPTION>
     <TITLE context="latex">$L/L_\odot$</TITLE>
  </FIELD>
</TABLE>
\end{verbatim}

In practice this extension would mean that, wherever a \elem{DESCRIPTION}
element is currently acceptable, a set of \elem{DESCRIPTION} and
\elem{TITLE} elements would become acceptable, each with an optional
\attr{context} additional attribute. The new \elem{TITLE} element
would have the role of expliciting the {\em column header} in a
field or parameter, or to supply a {\em caption} of a table or
a set of tables (resource) in addition to its description.

Providing descriptions in several languages would be another
obvious advantage of this extension.

\subsection{\texorpdfstring{A New {\tt XMLDATA} Serialization}
                           {A New XMLDATA Serialization}}
In order to facilitate the  use of standard XML query tools 
which usually require each parameter to have its own individual tag,
the \elem{XMLDATA} serialization introduces the designation of
each  \elem{FIELD} by a dedicated tag. An example could look like
the following:

\ifhtx\Beg{tabular}{\bg{LightCyan} CELLPADDING=5}{||l||}\fi
\begin{verbatim}
<TABLE name="Messier">
  <FIELD name="Number" ID="M" ucd="meta.record" datatype="int" >
    <DESCRIPTION>Messier Number</DESCRIPTION>
  </FIELD>
  <FIELD name="R.A.2000" ID="RA" ucd="pos.eq.ra;meta.main"
         unit="deg" datatype="float" width="5" precision="1" />
  <FIELD name="Dec.2000" ID="DE" ucd="pos.eq.dec;meta.main"
         unit="deg" datatype="float" width="5" precision="1" />
  <FIELD name="Name" ID="N" ucd="meta.id" datatype="char" arraysize="*">
    <DESCRIPTION>Common name used to designate the Messier object</DESCRIPTION>
  </FIELD>
  <FIELD ID="T" name="Classification" datatype="char" arraysize="10*" 
         ucd="src.class">
     <DESCRIPTION>Classification (galaxy, glubular cluster, etc)</DESCRIPTION>
  </FIELD>
  <DATA><XMLDATA>
    <TR>
      <M>3</M>
      <RA>205.5</RA>
      <DE>+28.4</DE>
      <N/>
      <T>Globular Cluster</T>
    </TR>
    <TR>
      <M>31</M>
      <RA>010.7</RA>
      <DE>+41.3</DE>
      <N>Andromeda Galaxy</N>
      <T>Galaxy</T>
    </TR>
  </XMLDATA></DATA>
</TABLE>
\end{verbatim}
\ifhtx\End{tabular}\fi
\par

\noindent The full document would need an XML-Schema definition of the tags
\elem{M}, \elem{RA}, \elem{DE}, \elem{N} and \elem{T}; these being
derived directly from the \attr{ID} attribute of the \elem{FIELD}
element, their definition can be generated automatically from the set of
\elem{FIELD} definitions.

\section{The VOTable V1.3 XML Schema}
\label{XML-schema}
The XML Schema of VOTable 1.3 is included here as a reference.
\revision{2}{
This schema includes a couple of extra optional attributes which are not
part of VOTable-1.2 ({\em ID} in TR and {\em encoding} in TD),
but proved to be useful to fix some problems encountered in the
usage of some code generators.
}{}
\ifhtx
\W{xsd}{http://www.ivoa.net/xml/VOTable/v1.3}
   {http://www.ivoa.net/xml/VOTable/v1.3}.
\par
\Beg{tabular}{\bg{LavenderBlush} CELLPADDING=10}{||l||}
\else
\bigskip
\small
\fi
\begin{verbatim}
<?xml version="1.0" encoding="UTF-8"?>
<!--W3C Schema for VOTable  = Virtual Observatory Tabular Format
.Version 1.0 : 15-Apr-2002
.Version 1.09: 23-Jan-2004 Version 1.09
.Version 1.09: 30-Jan-2004 Version 1.091
.Version 1.09: 22-Mar-2004 Version 1.092
.Version 1.094: 02-Jun-2004 GROUP does not contain FIELD
.Version 1.1 :  10-Jun-2004 remove the complexContent
.Version 1.11: GL: 23-May-2006 remove most root elements, use name= type= iso ref= structure
.Version 1.11: GL: 29-Aug-2006 review and added comments (prefixed by GL) 
              before sending to Francois Ochsenbein
.Version 1.12: FO: Preliminary Version 1.2
.Version 1.18: FO: Tested (jax) version 1.2
.Version 1.19: FO: Completed INFO attributes
.Version 1.20: FO: Added xtype; content-role is less restrictive (May2009)
.Version 1.20a: FO: PR-20090710 Cosmetics.
.Version 1.20b: FO: INFO does not accept sub-elements (2009-09-29)
.Version 1.20c: FO: elementFormDefault="qualified" to stay compatible with 1.1
.Version 1.3: MT: Added BINARY2 element
.Version 1.3: MT: Further relaxed LINK content-role type to token
-->
<xs:schema 
   xmlns:xs="http://www.w3.org/2001/XMLSchema" elementFormDefault="qualified"
   xmlns="http://www.ivoa.net/xml/VOTable/v1.3"
   targetNamespace="http://www.ivoa.net/xml/VOTable/v1.3" 
>
<xs:annotation><xs:documentation>
    VOTable is meant to serialize tabular documents in the
    context of Virtual Observatory applications. This schema
    corresponds to the VOTable document available from
    http://www.ivoa.net/Documents/latest/VOT.html
</xs:documentation></xs:annotation>

<!-- Here we define some interesting new datatypes:
     - anyTEXT   may have embedded XHTML (conforming HTML)
     - astroYear is an epoch in Besselian or Julian year, e.g. J2000
     - arrayDEF  specifies an array size e.g. 12x23x*
     - dataType  defines the acceptable datatypes
     - ucdType   defines the acceptable UCDs (UCD1+)
     - precType  defines the acceptable precisions
     - yesno     defines just the 2 alternatives
-->

<xs:complexType name="anyTEXT" mixed="true">
  <xs:sequence>
    <xs:any minOccurs="0" maxOccurs="unbounded" processContents="skip"/>
  </xs:sequence>
</xs:complexType>

<xs:simpleType  name="astroYear">
  <xs:restriction base="xs:token">
    <xs:pattern  value="[JB]?[0-9]+([.][0-9]*)?"/>
  </xs:restriction>
</xs:simpleType>

<xs:simpleType  name="ucdType">
  <xs:restriction base="xs:token">
    <xs:annotation><xs:documentation>
      Accept UCD1+
      Accept also old UCD1 (but not / + %) including SIAP convention (with :)
    </xs:documentation></xs:annotation>
    <xs:pattern  value="[A-Za-z0-9_.:;\-]*"/><!-- UCD1 use also / + % -->
  </xs:restriction>
</xs:simpleType>

<xs:simpleType  name="arrayDEF">
  <xs:restriction base="xs:token">
    <xs:pattern  value="([0-9]+x)*[0-9]*[*]?(s\W)?"/>
  </xs:restriction>
</xs:simpleType>

<xs:simpleType  name="encodingType">
  <xs:restriction base="xs:NMTOKEN">
    <xs:enumeration value="gzip"/>
    <xs:enumeration value="base64"/>
    <xs:enumeration value="dynamic"/>
    <xs:enumeration value="none"/>
  </xs:restriction>
</xs:simpleType>

<xs:simpleType name="dataType">
  <xs:restriction base="xs:NMTOKEN">
    <xs:enumeration value="boolean"/>
    <xs:enumeration value="bit"/>
    <xs:enumeration value="unsignedByte"/>
    <xs:enumeration value="short"/>
    <xs:enumeration value="int"/>
    <xs:enumeration value="long"/>
    <xs:enumeration value="char"/>
    <xs:enumeration value="unicodeChar"/>
    <xs:enumeration value="float"/>
    <xs:enumeration value="double"/>
    <xs:enumeration value="floatComplex"/>
    <xs:enumeration value="doubleComplex"/>
  </xs:restriction>
</xs:simpleType>

<xs:simpleType name="precType">
  <xs:restriction base="xs:token">
    <xs:pattern value="[EF]?[1-9][0-9]*"/>
  </xs:restriction>
</xs:simpleType>

<xs:simpleType name="yesno">
  <xs:restriction base="xs:NMTOKEN">
    <xs:enumeration value="yes"/>
    <xs:enumeration value="no"/>
  </xs:restriction>
</xs:simpleType>

  <xs:complexType name="Min">
    <xs:attribute name="value" type="xs:string" use="required"/>
    <xs:attribute name="inclusive" type="yesno" default="yes"/>
  </xs:complexType>
  <xs:complexType name="Max">
    <xs:attribute name="value" type="xs:string" use="required"/>
    <xs:attribute name="inclusive" type="yesno" default="yes"/>
  </xs:complexType>
  <xs:complexType name="Option">
    <xs:sequence>
      <xs:element name="OPTION" type="Option" minOccurs="0" maxOccurs="unbounded"/>
    </xs:sequence>
    <xs:attribute name="name" type="xs:token"/>
    <xs:attribute name="value" type="xs:string" use="required"/>
  </xs:complexType>
  
  <!-- VALUES expresses the values that can be taken by the data 
    in a column or by a parameter
  -->
  <xs:complexType name="Values">
    <xs:sequence>
      <xs:element name="MIN" type="Min" minOccurs="0"/>
      <xs:element name="MAX" type="Max" minOccurs="0"/>
      <xs:element name="OPTION" type="Option" minOccurs="0" maxOccurs="unbounded"/>
    </xs:sequence>
    <xs:attribute name="ID" type="xs:ID"/>
    <xs:attribute name="type" default="legal">
      <xs:simpleType>
        <xs:restriction base="xs:NMTOKEN">
          <xs:enumeration value="legal"/>
          <xs:enumeration value="actual"/>
        </xs:restriction>
      </xs:simpleType>
    </xs:attribute>
    <xs:attribute name="null" type="xs:token"/>
    <xs:attribute name="ref"  type="xs:IDREF"/>
    <!-- xs:attribute name="invalid" type="yesno" default="no"/ -->
  </xs:complexType>
  
  <!-- The LINK is a URL (href) or some other kind of reference (gref) -->
  <xs:complexType name="Link">
    <xs:annotation><xs:documentation> 
    content-role was previsouly restricted as: <![CDATA[
    <xs:attribute name="content-role">
      <xs:simpleType>
        <xs:restriction base="xs:NMTOKEN">
          <xs:enumeration value="query"/>
          <xs:enumeration value="hints"/>
          <xs:enumeration value="doc"/>
          <xs:enumeration value="location"/>
        </xs:restriction>
      </xs:simpleType>
    </xs:attribute>]]>; is now a token.
    </xs:documentation></xs:annotation>
    <xs:attribute name="ID" type="xs:ID"/>
    <xs:attribute name="content-role" type="xs:token"/>
    <xs:attribute name="content-type" type="xs:token"/>
    <xs:attribute name="title" type="xs:string"/>
    <xs:attribute name="value" type="xs:string"/>
    <xs:attribute name="href" type="xs:anyURI"/>
    <xs:attribute name="gref" type="xs:token"/><!-- Deprecated in V1.1 -->
    <xs:attribute name="action" type="xs:anyURI"/>
  </xs:complexType>
  
<!-- INFO is defined in Version 1.2 as a PARAM of String type 
<xs:complexType name="Info">
  <xs:complexContent>
    <xs:restriction base="Param">
      <xs:attribute name="unit" fixed=""/>
      <xs:attribute name="datatype" fixed="char"/>
      <xs:attribute name="arraysize" fixed="*"/>
    </xs:restriction>
  </xs:complexContent>
</xs:complexType>
 -or- as a full definition:
<xs:complexType name="Info">
  <xs:sequence> 
  <xs:element name="DESCRIPTION" type="anyTEXT" minOccurs="0"/>
    <xs:element name="VALUES" type="Values" minOccurs="0"/>
    <xs:element name="LINK" type="Link" minOccurs="0" maxOccurs="unbounded"/> 
  </xs:sequence>
  <xs:attribute name="name" type="xs:token" use="required"/>
  <xs:attribute name="value" type="xs:string" use="required"/>
  <xs:attribute name="ID" type="xs:ID"/>
  <xs:attribute name="unit" type="xs:token"/>
  <xs:attribute name="xtype" type="xs:token"/>
  <xs:attribute name="ref" type="xs:IDREF"/>
  <xs:attribute name="ucd" type="ucdType"/>
  <xs:attribute name="utype" type="xs:string"/>
</xs:complexType>
-->
<!-- No sub-element is accepted in INFO for backward compatibility -->
<xs:complexType name="Info">
  <xs:simpleContent>
    <xs:extension base="xs:string">
      <xs:attribute name="ID" type="xs:ID"/>
      <xs:attribute name="name"  type="xs:token" use="required"/>
      <xs:attribute name="value" type="xs:string" use="required"/>
      <xs:attribute name="unit"  type="xs:token"/>
      <xs:attribute name="xtype" type="xs:token"/>
      <xs:attribute name="ref"   type="xs:IDREF"/>
      <xs:attribute name="ucd"   type="ucdType"/>
      <xs:attribute name="utype" type="xs:string"/>
    </xs:extension>
  </xs:simpleContent>
</xs:complexType>

<!-- Expresses the coordinate system we are using --><!-- Deprecated V1.2 -->
<xs:complexType name="CoordinateSystem">
  <xs:annotation><xs:documentation>
    Deprecated in Version 1.2
  </xs:documentation></xs:annotation>
  <xs:simpleContent>
    <xs:extension base="xs:string">
      <xs:attribute name="ID" type="xs:ID" use="required"/>
      <xs:attribute name="equinox" type="astroYear"/>
      <xs:attribute name="epoch" type="astroYear"/>
      <xs:attribute name="system" default="eq_FK5">
        <xs:simpleType>
          <xs:restriction base="xs:NMTOKEN">
            <xs:enumeration value="eq_FK4"/>
            <xs:enumeration value="eq_FK5"/>
            <xs:enumeration value="ICRS"/>
            <xs:enumeration value="ecl_FK4"/>
            <xs:enumeration value="ecl_FK5"/>
            <xs:enumeration value="galactic"/>
            <xs:enumeration value="supergalactic"/>
            <xs:enumeration value="xy"/>
            <xs:enumeration value="barycentric"/>
            <xs:enumeration value="geo_app"/>
          </xs:restriction>
        </xs:simpleType>
      </xs:attribute>
    </xs:extension>
  </xs:simpleContent>
</xs:complexType>

<xs:complexType name="Definitions">
  <xs:annotation><xs:documentation>
    Deprecated in Version 1.1
  </xs:documentation></xs:annotation>
  <xs:choice minOccurs="0" maxOccurs="unbounded">
    <xs:element name="COOSYS" type="CoordinateSystem"/><!-- Deprecated in V1.2 -->
    <xs:element name="PARAM" type="Param"/>
  </xs:choice>
</xs:complexType>

<!-- FIELD is the definition of what is in a column of the table -->
<xs:complexType name="Field">
  <xs:sequence> <!-- minOccurs="0" maxOccurs="unbounded" -->
    <xs:element name="DESCRIPTION" type="anyTEXT" minOccurs="0"/>
    <xs:element name="VALUES" type="Values" minOccurs="0"/> <!-- maxOccurs="2" -->
    <xs:element name="LINK" type="Link" minOccurs="0" maxOccurs="unbounded"/>
  </xs:sequence>
  <xs:attribute name="ID" type="xs:ID"/>
  <xs:attribute name="unit" type="xs:token"/>
  <xs:attribute name="datatype" type="dataType" use="required"/>
  <xs:attribute name="precision" type="precType"/>
  <xs:attribute name="width" type="xs:positiveInteger"/>
  <xs:attribute name="xtype" type="xs:token"/>
  <xs:attribute name="ref" type="xs:IDREF"/>
  <xs:attribute name="name" type="xs:token" use="required"/>
  <xs:attribute name="ucd" type="ucdType"/>
  <xs:attribute name="utype" type="xs:string"/>
  <xs:attribute name="arraysize" type="xs:string"/>
    <!-- GL: is the next deprecated element remaining 
        (is not in PARAM, but will in new model be inherited) 
    -->
  <xs:attribute name="type">
    <!-- type is not in the Version 1.1, but is kept for
         backward compatibility purposes
    -->
    <xs:simpleType>
      <xs:restriction base="xs:NMTOKEN">
        <xs:enumeration value="hidden"/>
        <xs:enumeration value="no_query"/>
        <xs:enumeration value="trigger"/>
        <xs:enumeration value="location"/>
      </xs:restriction>
    </xs:simpleType>
  </xs:attribute>
</xs:complexType>


<!-- A PARAM is similar to a FIELD, but it also has a "value" attribute -->
<!--  GL: implemented here as a subtype as suggested we do in Kyoto. -->
<xs:complexType name="Param">
  <xs:complexContent>
    <xs:extension base="Field">
      <xs:attribute name="value" type="xs:string" use="required"/>
    </xs:extension>
  </xs:complexContent>
</xs:complexType>


<!-- GROUP groups columns; may include descriptions, fields/params/groups -->
<xs:complexType name="Group">
  <xs:sequence>
    <xs:element name="DESCRIPTION" type="anyTEXT" minOccurs="0"/>
<!--  GL I guess I can understand the next choice element as one may (?) 
      really want to group fields and params and groups in a particular order.
-->    
    <xs:choice minOccurs="0" maxOccurs="unbounded">
      <xs:element name="FIELDref" type="FieldRef"/> 
      <xs:element name="PARAMref" type="ParamRef"/> 
      <xs:element name="PARAM" type="Param"/> 
      <xs:element name="GROUP" type="Group"/> 
      <!-- GL a GroupRef could remove recursion -->
    </xs:choice>
  </xs:sequence>
  <xs:attribute name="ID"   type="xs:ID"/>
  <xs:attribute name="name" type="xs:token"/>
  <xs:attribute name="ref"  type="xs:IDREF"/>
  <xs:attribute name="ucd"  type="ucdType"/>
  <xs:attribute name="utype" type="xs:string"/>
</xs:complexType>

<!-- FIELDref and PARAMref are references to FIELD or PARAM defined
     in the parent TABLE or RESOURCE -->
<!-- GL This can not be enforced in XML Schema, so why not IDREF in <Group> ?
     In particular if the UCD and utype attributes will NOT be added -->
<xs:complexType name="FieldRef">
  <xs:attribute name="ref" type="xs:IDREF" use="required"/>
  <xs:attribute name="ucd"  type="ucdType"/>
  <xs:attribute name="utype" type="xs:string"/>
</xs:complexType>

<xs:complexType name="ParamRef">
  <xs:attribute name="ref" type="xs:IDREF" use="required"/>
  <xs:attribute name="ucd"  type="ucdType"/>
  <xs:attribute name="utype" type="xs:string"/>
</xs:complexType>

<!-- DATA is the actual table data, in one of three formats -->
<!-- 
  GL in Kyoto we discussed the option of having the specific Data items 
  be subtypes of Data:
-->
<!-- 
<xs:complexType name="Data" abstract="true"/>

<xs:complexType name="TableData">
  <xs:complexContent>
    <xs:extension base="Data">
     ... etc
    </xs:extension>
  </xs:complexContent>
</xs:complexType>
 -->
<xs:complexType name="Data">
  <xs:annotation><xs:documentation>
    Added in Version 1.2: INFO for diagnostics
  </xs:documentation></xs:annotation>
  <xs:sequence>
    <xs:choice>
      <xs:element name="TABLEDATA" type="TableData"/>
      <xs:element name="BINARY" type="Binary"/>
      <xs:element name="BINARY2" type="Binary2"/>
      <xs:element name="FITS" type="FITS"/>
    </xs:choice>
    <xs:element name="INFO" type="Info" minOccurs="0" maxOccurs="unbounded"/>
  </xs:sequence>
</xs:complexType>

<!-- Pure XML data -->
<xs:complexType name="TableData">
  <xs:sequence>
    <xs:element name="TR" type="Tr" minOccurs="0" maxOccurs="unbounded"/>
  </xs:sequence>
</xs:complexType>

<xs:complexType name="Td">
  <xs:simpleContent>
    <xs:extension base="xs:string">
      <!-- xs:attribute name="ref" type="xs:IDREF"/ -->
      <xs:annotation><xs:documentation>
          The 'encoding' attribute is added here to avoid
          problems of code generators which do not properly
          interpret the TR/TD structures.
          'encoding' was chosen because it appears in
          appendix A.5
      </xs:documentation></xs:annotation>
      <xs:attribute name="encoding" type="encodingType"/>
    </xs:extension>
  </xs:simpleContent>
</xs:complexType>

<xs:complexType name="Tr">
  <xs:annotation><xs:documentation>
    The ID attribute is added here to the TR tag to avoid 
    problems of code generators which do not properly 
    interpret the TR/TD structures
  </xs:documentation></xs:annotation>
  <xs:sequence>
    <xs:element name="TD" type="Td" maxOccurs="unbounded"/>
  </xs:sequence>
  <xs:attribute name="ID" type="xs:ID"/>
</xs:complexType>

<!-- FITS file, perhaps with specification of which extension to seek to -->
<xs:complexType name="FITS">
  <xs:sequence>
    <xs:element name="STREAM" type="Stream"/>
  </xs:sequence>
  <xs:attribute name="extnum" type="xs:positiveInteger"/>
</xs:complexType>

<!-- BINARY data format -->
<xs:complexType name="Binary">
  <xs:sequence>
    <xs:element name="STREAM" type="Stream"/>
  </xs:sequence>
</xs:complexType>

<!-- BINARY2 data format -->
<xs:complexType name="Binary2">
  <xs:sequence>
    <xs:element name="STREAM" type="Stream"/>
  </xs:sequence>
</xs:complexType>

<!-- STREAM can be local or remote, encoded or not -->
<xs:complexType name="Stream">
  <xs:simpleContent>
    <xs:extension base="xs:string">
      <xs:attribute name="type" default="locator">
        <xs:simpleType>
          <xs:restriction base="xs:NMTOKEN">
            <xs:enumeration value="locator"/>
            <xs:enumeration value="other"/>
          </xs:restriction>
        </xs:simpleType>
      </xs:attribute>
      <xs:attribute name="href" type="xs:anyURI"/>
      <xs:attribute name="actuate" default="onRequest">
        <xs:simpleType>
          <xs:restriction base="xs:NMTOKEN">
            <xs:enumeration value="onLoad"/>
            <xs:enumeration value="onRequest"/>
            <xs:enumeration value="other"/>
            <xs:enumeration value="none"/>
          </xs:restriction>
        </xs:simpleType>
      </xs:attribute>
      <xs:attribute name="encoding" type="encodingType" default="none"/>
      <xs:attribute name="expires" type="xs:dateTime"/>
      <xs:attribute name="rights" type="xs:token"/>
    </xs:extension>
  </xs:simpleContent>
</xs:complexType>

<!-- A TABLE is a sequence of FIELD/PARAMs and LINKS and DESCRIPTION, 
     possibly followed by a DATA section 
-->
<xs:complexType name="Table">
  <xs:annotation><xs:documentation>
    Added in Version 1.2: INFO for diagnostics
  </xs:documentation></xs:annotation>
  <xs:sequence>
    <xs:element name="DESCRIPTION" type="anyTEXT" minOccurs="0"/>
<!-- GL: why a choice iso for example -->
<!-- 
      <xs:element name="PARAM" type="Param" minOccurs="0" maxOccurs="unbounded"/>
      <xs:element name="FIELD" type="Field" minOccurs="0" maxOccurs="unbounded"/>
      <xs:element name="GROUP" type="Group" minOccurs="0" maxOccurs="unbounded"/>
-->
<!-- 
  This could also enforce groups to be defined after the fields and params 
  to which they must have a reference, which is somewhat more logical
-->
    <!-- Added Version 1.2: -->
    <xs:element name="INFO" type="Info" minOccurs="0" maxOccurs="unbounded"/> 
    <!-- An empty table without any FIELD/PARAM should not be acceptable -->
    <xs:choice minOccurs="1" maxOccurs="unbounded"> 
      <xs:element name="FIELD" type="Field"/>
      <xs:element name="PARAM" type="Param"/>
      <xs:element name="GROUP" type="Group"/>
    </xs:choice>
    <xs:element name="LINK" type="Link" minOccurs="0" maxOccurs="unbounded"/>
    <!-- This would allow several DATA parts in a table (future extension?)
    <xs:sequence minOccurs="0" maxOccurs="unbounded">  
      <xs:element name="DATA" type="Data"/>
      <xs:element name="INFO" type="Info" minOccurs="0" maxOccurs="unbounded"/>
    </xs:sequence>
    -->
    <xs:element name="DATA" type="Data" minOccurs="0"/>
    <xs:element name="INFO" type="Info" minOccurs="0" maxOccurs="unbounded"/>
  </xs:sequence>
  <xs:attribute name="ID"   type="xs:ID"/>
  <xs:attribute name="name" type="xs:token"/>
  <xs:attribute name="ref"  type="xs:IDREF"/>
  <xs:attribute name="ucd"  type="ucdType"/>
  <xs:attribute name="utype" type="xs:string"/>
  <xs:attribute name="nrows" type="xs:nonNegativeInteger"/>
</xs:complexType>

<!-- RESOURCES can contain DESCRIPTION, (INFO|PARAM|COSYS), LINK, TABLEs -->
<xs:complexType name="Resource">
  <xs:annotation><xs:documentation>
     Added in Version 1.2: INFO for diagnostics in several places
  </xs:documentation></xs:annotation>
  <xs:sequence>
    <xs:element name="DESCRIPTION" type="anyTEXT" minOccurs="0"/>
    <xs:element name="INFO" type="Info" minOccurs="0" maxOccurs="unbounded"/>
    <xs:choice minOccurs="0" maxOccurs="unbounded">
      <xs:element name="COOSYS" type="CoordinateSystem"/><!-- Deprecated in V1.2 -->
      <xs:element name="GROUP" type="Group" />
      <xs:element name="PARAM" type="Param" />
    </xs:choice>
    <xs:sequence minOccurs="0" maxOccurs="unbounded">
      <xs:element name="LINK" type="Link" minOccurs="0" maxOccurs="unbounded"/>
      <xs:choice>
        <xs:element name="TABLE" type="Table" />
        <xs:element name="RESOURCE" type="Resource" />
      </xs:choice>
      <xs:element name="INFO" type="Info" minOccurs="0" maxOccurs="unbounded"/>
    </xs:sequence>
    <!-- Suggested Doug Tody, to include new RESOURCE types -->
    <xs:any namespace="##other" processContents="lax" minOccurs="0" maxOccurs="unbounded"/>
  </xs:sequence>
  <xs:attribute name="name" type="xs:token"/>
  <xs:attribute name="ID"   type="xs:ID"/>
  <xs:attribute name="utype" type="xs:string"/>
  <xs:attribute name="type" default="results">
    <xs:simpleType>
      <xs:restriction base="xs:NMTOKEN">
        <xs:enumeration value="results"/>
        <xs:enumeration value="meta"/>
      </xs:restriction>
    </xs:simpleType>
  </xs:attribute>
  <!-- Suggested Doug Tody, to include new RESOURCE attributes -->
  <xs:anyAttribute namespace="##other" processContents="lax"/>
</xs:complexType>

<!-- VOTable is the root element -->
<xs:element name="VOTABLE">
<xs:complexType>
  <xs:sequence>
    <xs:element name="DESCRIPTION" type="anyTEXT" minOccurs="0"/>
    <xs:element name="DEFINITIONS" type="Definitions" minOccurs="0"/><!-- Deprecated -->
    <xs:choice minOccurs="0" maxOccurs="unbounded">
      <xs:element name="COOSYS" type="CoordinateSystem"/><!-- Deprecated in V1.2 -->
      <xs:element name="GROUP" type="Group" />
      <xs:element name="PARAM" type="Param" />
      <xs:element name="INFO" type="Info" minOccurs="0" maxOccurs="unbounded"/>
    </xs:choice>
    <xs:element name="RESOURCE" type="Resource" minOccurs="1" maxOccurs="unbounded"/>
    <xs:element name="INFO" type="Info" minOccurs="0" maxOccurs="unbounded"/>
  </xs:sequence>
  <xs:attribute name="ID" type="xs:ID"/>
  <xs:attribute name="version">
     <xs:simpleType>
       <xs:restriction base="xs:NMTOKEN">
         <xs:enumeration value="1.3"/>
       </xs:restriction>
     </xs:simpleType>
   </xs:attribute>
</xs:complexType>
</xs:element>

</xs:schema>
\end{verbatim}
\ifhtx\End{tabular}\else
\normalsize
\fi
\end{document}